%% file: 000_main.tex
\documentclass{amsart}
\usepackage[utf8]{inputenc}

\input{000_setup}

\title[App: Poncelet Triangle Loci]{An App for the Discovery of\\Properties of Poncelet Triangles\vspace{-1em}}
\author[I. Darlan]{Iverton Darlan} 
\thanks{I. Darlan, Federal Univ. of Rio de Janeiro, Brazil.
\texttt{iverton1996@gmail.com}
}
\author[D. Reznik]{Dan Reznik}
\thanks{D. Reznik$^*$, Data Science Consulting Ltd., Rio de Janeiro, Brazil. \texttt{dreznik@gmail.com}}
\date{September, 2020}

\begin{document}

\maketitle

\vspace{-2em}
\input{005_abstract}

\tableofcontents

\input{010_intro}
\input{040_secs}


\appendix
\input{910_app_poncelet}

\bibliographystyle{maa}
\bibliography{refs,refs_rgk,refs_rgk_private,refs_rgk_media}

\end{document}

%% file: 000_setup.tex
\usepackage{graphics}
\usepackage{thmtools}
\usepackage{wasysym}
\usepackage[T1]{fontenc}    

\usepackage{amsthm}
\usepackage{amsbsy,amsmath,amssymb,amscd,amsfonts}
\usepackage[pagebackref=true]{hyperref}

\usepackage{graphicx,float,latexsym,color}
\usepackage[font={scriptsize,it}]{caption}
\usepackage{subcaption}

\usepackage{makecell}

\usepackage[dvipsnames]{xcolor}

\newtheorem*{theorem*}{Theorem}

\theoremstyle{remark}

\theoremstyle{definition}

\hypersetup{
    pdftoolbar=true,        
    pdfmenubar=true,        
    pdffitwindow=false,     
    pdfstartview={FitH},    
    colorlinks=true,       
    linkcolor=OliveGreen,          
    citecolor=blue,        
    filecolor=black,      
    urlcolor=red           
}

\usepackage[nameinlink,capitalize,noabbrev]{cleveref}

\usepackage{lineno}

\usepackage{comment}
\usepackage{microtype}

\graphicspath{{pics/}}
\newcommand{\Tm}{\mathcal{T}}
\newcommand{\Lm}{\mathcal{L}}
\newcommand{\Cm}{\mathcal{C}}
\newcommand{\E}{\mathcal{E}}

%% file: 005_abstract.tex
\begin{abstract}
We describe a newly-developed, free, browser-based application, for the interactive exploration of the dynamic geometry of Poncelet families of triangles. The main focus is on responsive display of the beauteous loci of centers of such families, refreshing them smoothly upon any changes in simulation parameters. The app informs the user when curves swept are conics and reports if certain metric quantities are conserved. Live simulations can be easily shared via a URL. A list of more than 400 pre-made experiments is included which can be regarded as conjectures and/or exercises. Millions of experiment combinations are possible.
\end{abstract}



%% file: 010_intro.tex
\section{Introduction}
\label{sec:intro}

\begin{figure}
    \centering
    \includegraphics[width=\textwidth]{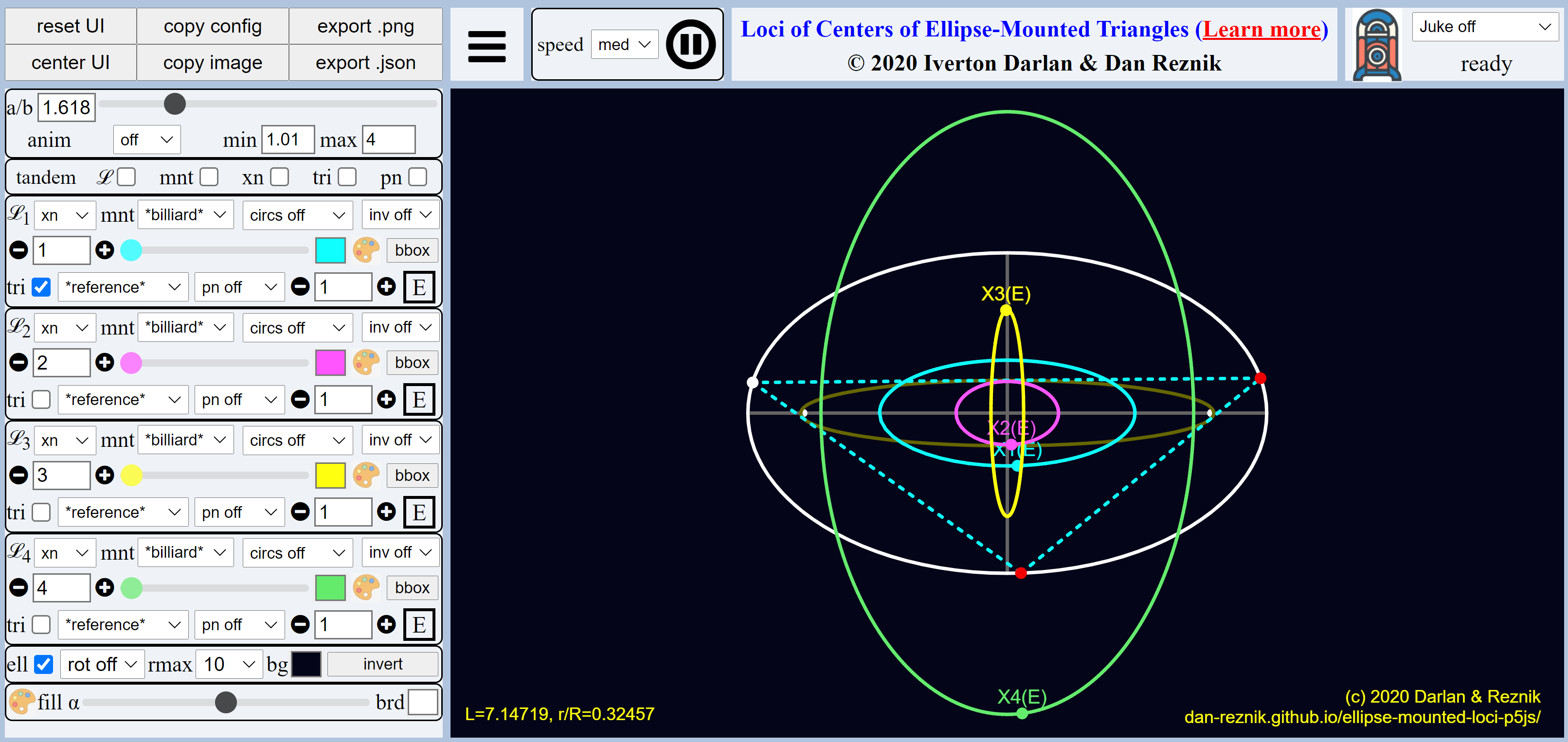}
    \caption{Locus Visualization app to explore 3-periodic families. Shown are the loci of $X_k$, $k=$1,2,3,4, over billiard 3-periodics. The ``(E)'' suffix indicated they are numerically ellipses. \href{https://bit.ly/3yV8caF}{Live}; Also see our tutorial \href{https://bit.ly/3iCKHxn}{playlist}.}
    \label{fig:09-screenshot}
\end{figure}

Referring to \cref{fig:09-screenshot}, we describe a web-based app (live \href{https://dan-reznik.github.io/ellipse-mounted-loci-p5js}{here}) for the interactive, shareable discovery of properties the dynamic geometry of ``Poncelet'' triangle families, in the spirit of \cite{davis1995-experimental}. Recall these are triangles inscribed in a conic while simultaneously circumscribing another conic. In our case, said conics are a pair of nested ellipses. \cref{app:poncelet} reviews Poncelet's porism.

\begin{figure}
    \centering
    \includegraphics[trim=0 0 75 0,clip,width=\textwidth]{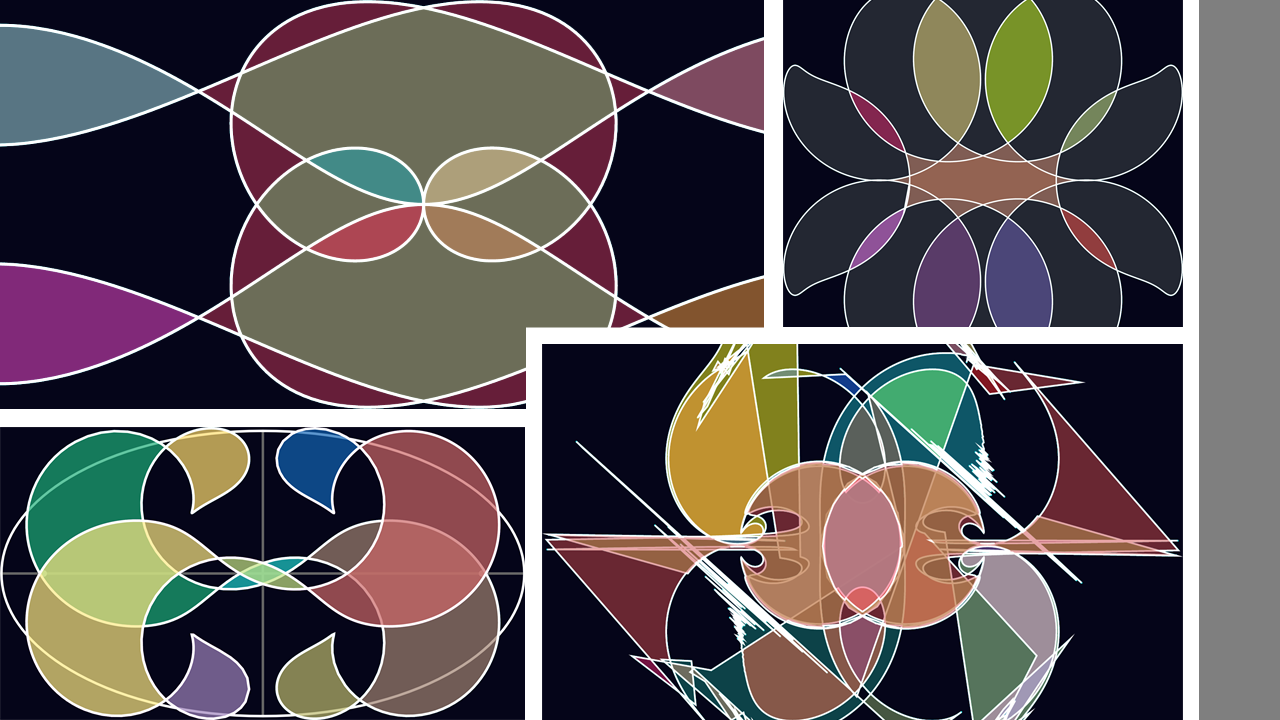}
    \caption{Four examples of the kinds of color-filled loci which can be produced with the app. \href{https://bit.ly/34sO8Px}{Gallery} and \href{https://youtu.be/l-O5UT8tpuw}{Video}}
    \label{fig:09-showcase}
\end{figure}

The main thrust is the interactive, real-time rendering of loci of notable points of a triangle, such as the incenter barycenter, etc. As shown in \cref{fig:09-showcase}, one initial interest was the aesthetics of loci which can be produced.

The app supports the first 1000 triangle centers in \cite{etc} are supported. We also want all experiments to be universally observable and easily shared, namely, via simple URLs. Another goal is to enable the user to quickly and visually assess the behavior of dozens of related experiments.

We built this app out of our frustration with certain go-to tools. Take Mathematica  \cite{mathematica_v10} for example: its programmatic interface is excellent for prototyping generic geometric simulations (we have written hundreds of them) and doing serious research. However locus computation is slow loci and sharing experiments is cumbersome. Geogebra \cite{geogebra2013}, on the other hand, is excellent for web sharing, education, but requires much manual construction. Its programmatic interface is cumbersome and specialized. Rendering of loci is slow, manual, and of low precision. In both cases, setting up Poncelet-related simulations requires much rework, so we felt a specialized app was in order.

\begin{figure}
    \centering
    \includegraphics[width=\textwidth]{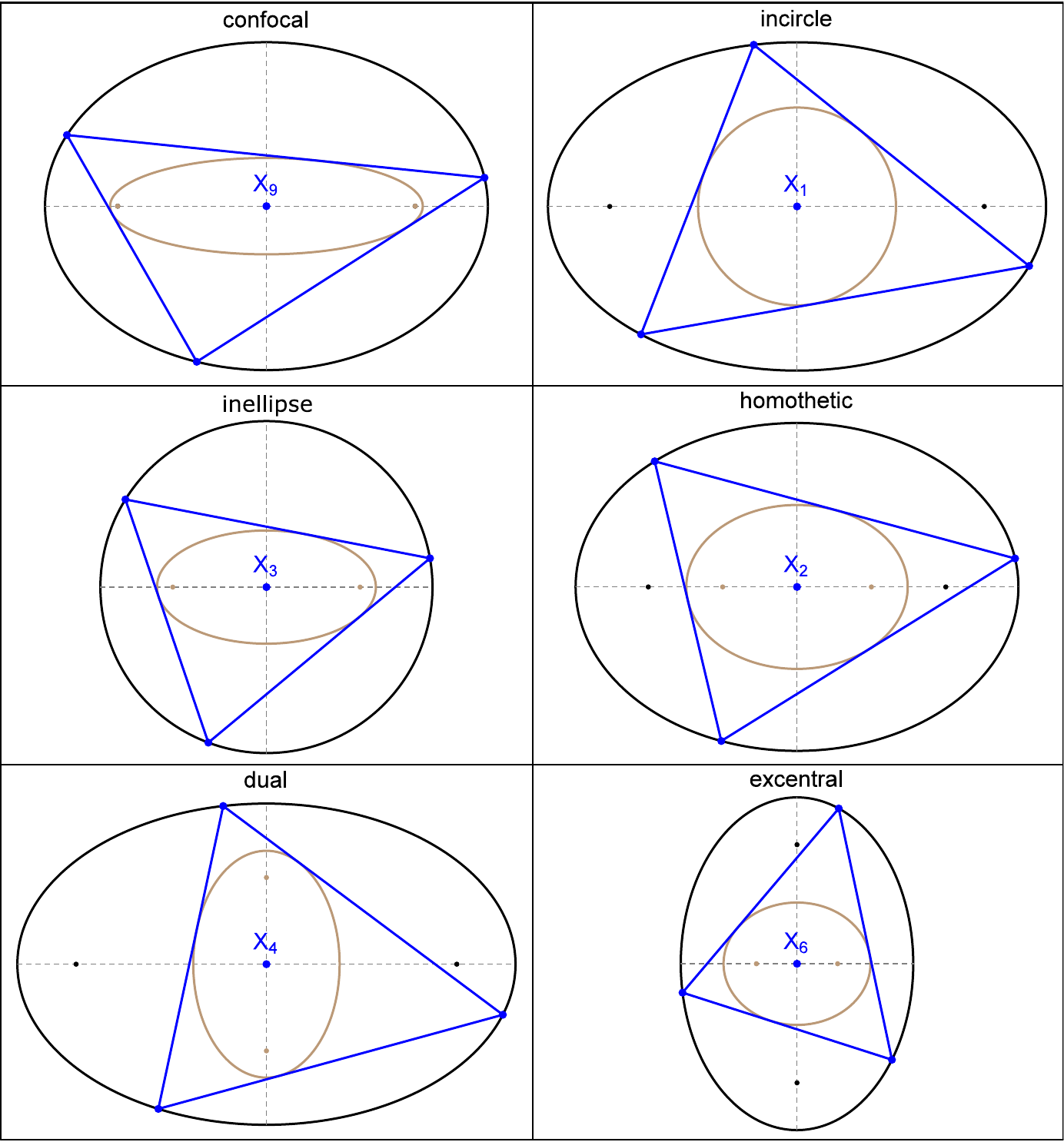}
    \caption{The confocal family is shown at the top left. Also shown are 5 other ``famous'' concentric families. \href{https://youtu.be/14TQ5WlZxUw}{Video}}
    \label{fig:six-caps}
\end{figure}

\begin{figure}
\centering
\includegraphics[width=\textwidth]{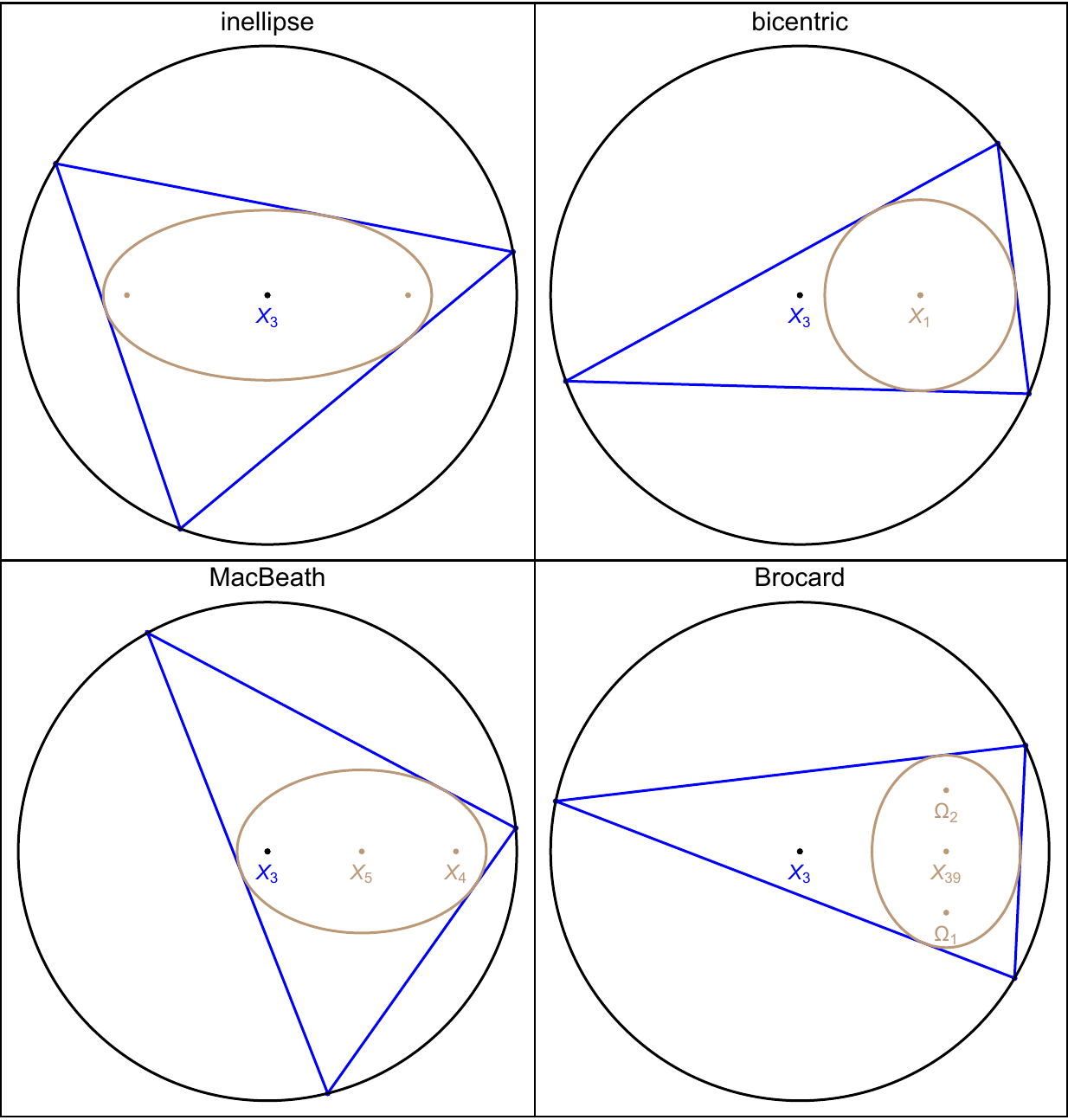}
\caption{Four circle-inscribed families supported: \textbf{inellipse} (top-left): ``inellipse'', caustic is a concentric inellipse (appears in \cref{fig:six-caps}); \textbf{bicentric} (top-right): Chapple's porism, caustic is the fixed incircle; \textbf{macbeath} (bottom-left), excentrals to the bicentric family, caustic is the MacBeath inconic \cite{mw}, with foci on $X_3$ and $X_4$; \textbf{brocard} (bottom-right): also known as the Brocard porism \cite{bradley2011-brocard}, or the ``harmonic family'', caustic is the Brocard inellipse \cite{mw}, with foci on the two Brocard points $\Omega_1,\Omega_2$.}
\label{fig:por-broc}
\end{figure}

\begin{figure}
\centering
\includegraphics[width=\textwidth]{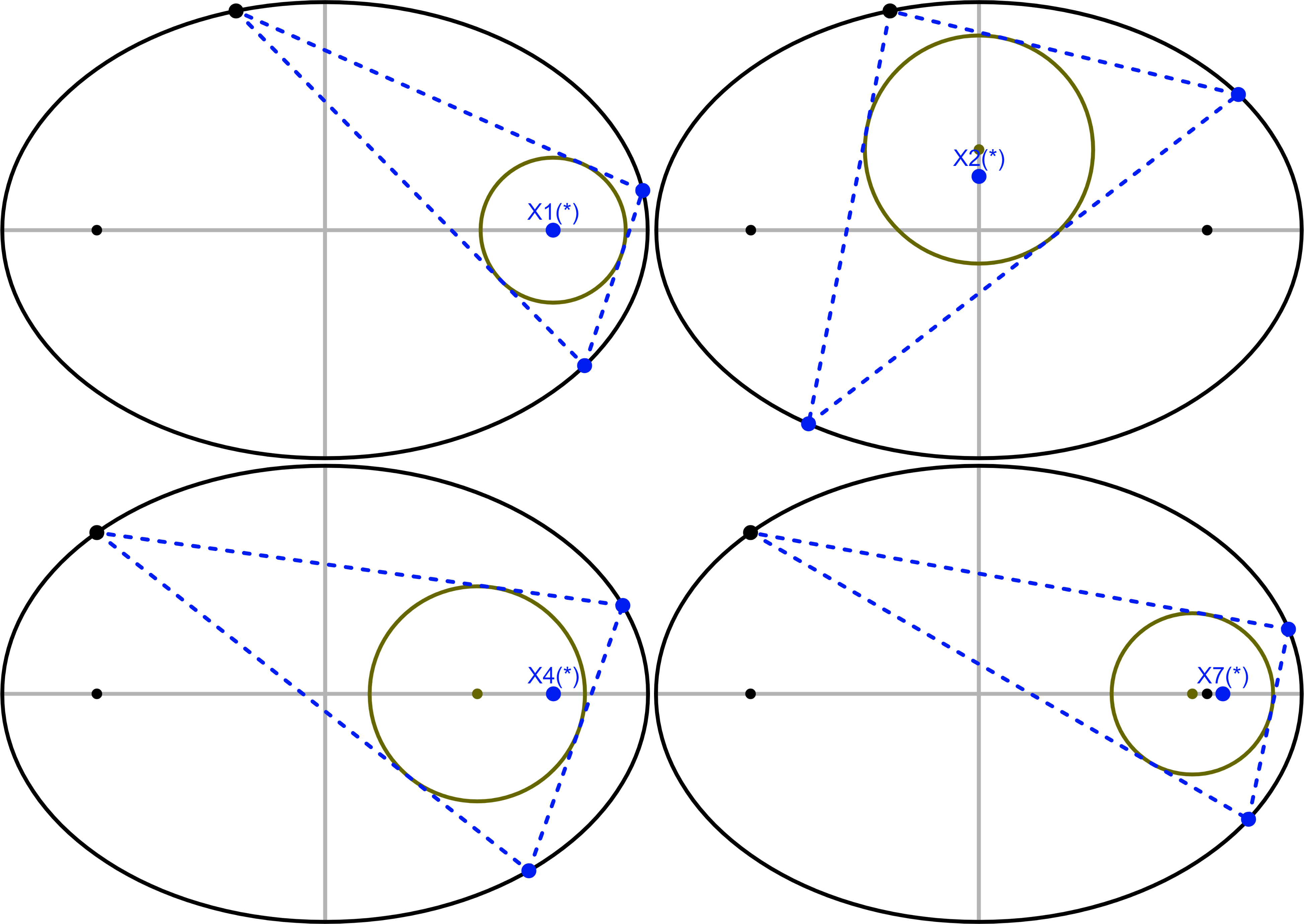}
\caption{Four ellipse-inscribed families with non-concentric, circular caustics: \textbf{top-left}: stationary incenter $X_1$ (at a focus); \textbf{top-right}: stationary barycenter $X_2$ (and Nagel point $X_8$ at center); \textbf{bottom-left}: stationary orthocenter $X_4$ (at a focus); \textbf{bottom-right}: [to do] stationary Gergonne point $X_7$. Caustic centers and radii kindly derived by Ronaldo Garcia \cite{garcia2024-private}.}
\label{fig:circ-caustic}
\end{figure}

The app supports the following groups of Poncelet families:

\begin{enumerate}
\item \cref{fig:six-caps}: six ellipse-inscribed families with concentric, axis-parallel caustics.
\item \cref{fig:por-broc}: four non-concentric, circle-inscribed families.
\item \cref{fig:circ-caustic}: four ellipse-inscribed, non-concentric, circular caustic families.
\end{enumerate}

An additional dozen-odd non-Ponceletian families are supported where two vertices remain stationary while a third one rides on a curve related to a base ellipse.

\begin{figure}
\centering
\includegraphics[width=\textwidth]{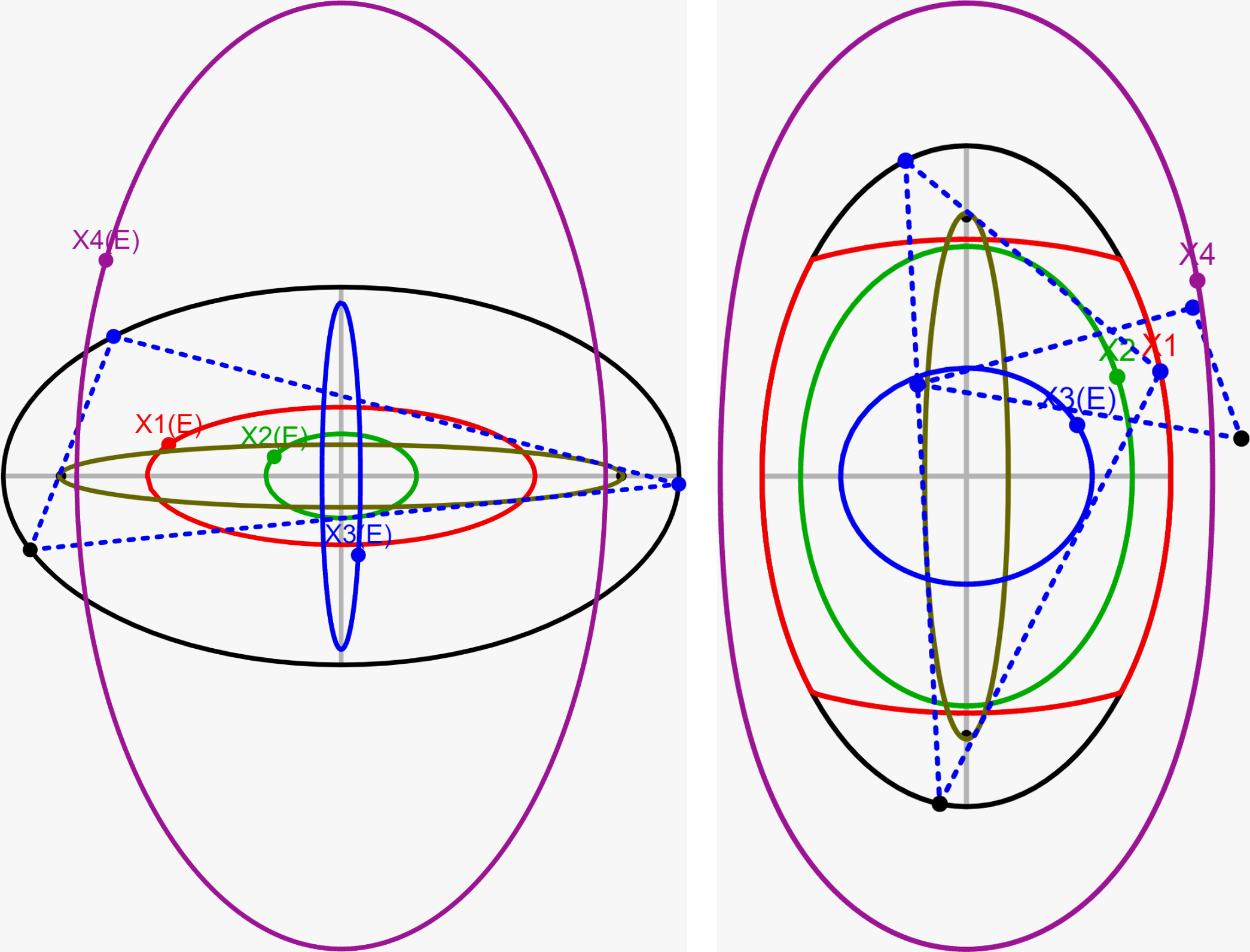}
\caption{\texttt{Left}: simultaneous disply of the loci of incenter $X_1$, barycenter $X_2$, circumcenter $X_3$, orthocenter $X_4$ over triangles in a confocal pair of ellipses (also known as elliptic billiard 3-periodics). An ``(E)'' next to each label indicates that numerically, each displayed locus is an ellipse; \href{https://bit.ly/3ACTkhh}{live}. \texttt{Right}: loci of the same centers for the {\em orthic} triangles (smaller dashed blue) derived from the confocal family (rotated $90^\circ$ for image fitting). Notice only the circumcenter $X_3$ is still an ellipse. The locus of the incenter $X_1$ has four kinks (it is actually piecewise elliptic); \href{https://bit.ly/2XEFhJB}{live}.}
\label{fig:sample-loci}
\end{figure}

\cref{fig:sample-loci} provides an example of locus visualization over a certain family for either a reference triangle or some derivative thereof (e.g., orthic, excentral, etc.). A typical workflow involves: (i) selecting a triangle family, (ii) selecting the triangle centers whose loci will be rendered (up to four can be displayed simultaneously), and (iii) whether the actual Poncelet or a derived triangle (such as the orthic, excentral, etc.) should be used as a basis for locus computation.

The app supports more than 50 derived triangles. With 6 triangle families and 1000 triangle centers, this amounts to some 300k experimental possibilities, each smoothly modulated by a continuous choice of aspect ratio of the outer ellipse. In fact many other possible discrete parameters amount to 10s of millions of possible combinations.

\begin{figure}
    \centering
    \includegraphics[width=\textwidth]{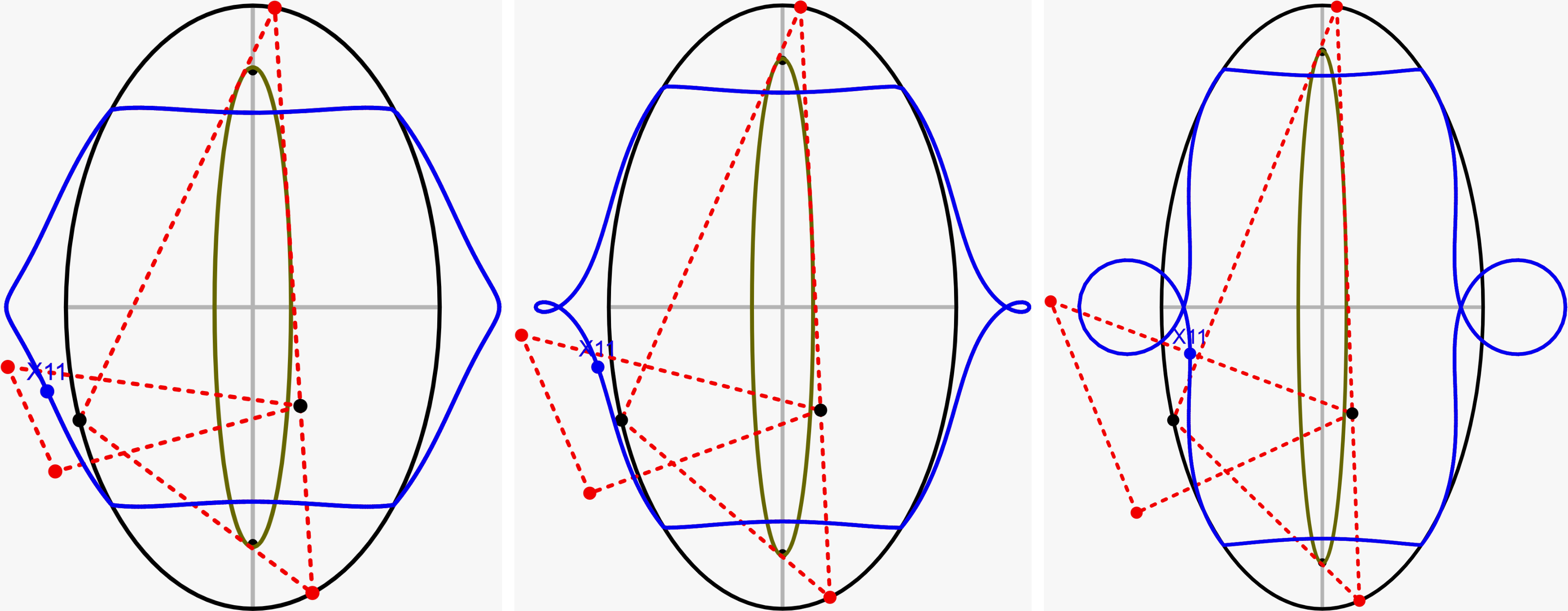}
    \caption{Observing transitions in the topology of tghe locus of the Feuerbach point $X_11$ of the orthic triangle (smaller dashed red) as the aspect ratio of the outer ellipse in the confocal pair is smoothly altered; \href{https://bit.ly/2Zg1caK}{live}.}
    \label{fig:orthic}
\end{figure}

Crucially, as the user changes any experimental parameter, loci are smoothly recomputed and re-displayed. As illustrated in \cref{fig:orthic}, this  allows one to observe curious changes in locus topology and pinpoint any critical points. 

\subsection*{Locus type and conservations} the app also informs the user if (i) the curve swept by a triangle center is a conic, point, or line, and (ii) if any metric quantities are conserved, out of certain presets (perimeter, areas, sums of trigonometric functions of angles, ratios, etc.). As shown in \cref{tab:n3-conc-families}, this tool and previous work with Mathematica has pointed us to many new curious conserved quantities manifested by each individual family, some proved, some simply reported in previous publications, e.g.,  \cite{akopyan2020-invariants,caliz2020-area-product,bialy-maxim-2018,reznik2021-fifty-invariants}.
 
\begin{table}
\centering
\begin{tabular}{|r|c|c|c|c|}
\hline
Family & Fixed & CAP & CI & Conserves\\
\hline
Confocal & $X_9$ & Y & N & \makecell[cc]{$L, J, r/R, \sum\cos\theta_i, \kappa_i^{2/3}$}  \\
\hline
Incircle & $X_1$ & Y & N & $R, \sum\cos\theta_i$ \\
\hline
\makecell[rc]{Confocal\\Excentrals} & $X_6$ & Y & N & \makecell[cc]{$A'/A, \prod\cos\theta_i', \sum{(s_i')^2}/\prod{s_i'}$}  \\
\hline
Homothetic & $X_2$ & Y & N & \makecell[cc]{$A, \sum{s_i^2}, \omega, \sum\cot\theta_i, \kappa_i^{-2/3}, \kappa_i^{-4/3}$}  \\
\hline
Dual & $X_4$ & Y & N & $\rho^2=4R^2-(\sum{s_i^2})/2$ \\
\hline
\hline
Inellipse & $X_3$ & Y & Y & \makecell[cc]{$\sum{s_i^2}, \prod\cos\theta_i, r_h, R_h, (\kappa_i')^{2/3}, \rho^2$} \\
\hline
Bicentric & $X_1,X_3$ & N & Y & \makecell[cc]{ $r/R$, $\sum\cos\theta_i$} \\
\hline
Brocard & $X_3,X_6$ & N & Y & \makecell[cc]{ $\sum\cot\theta_i, \sum{s_i^2}/A$} \\
\hline
MacBeath & $X_3,X_4$ & N & Y & $\sum\cos(2\theta_i), \prod\cos(\theta_i), \sum{s_i}^2, \rho^2$ \\
\hline
\hline
iso-$X_1$ & $X_1$ & N & N & $\sum\sin(\theta_i/2)$ \\
\hline
iso-$X_2$ & $X_k,k=1,2,8,10,...$ & N & N & n/a \\
\hline
iso-$X_4$ & $X_k,k=1,4,7952,18283,...$ & N & N & $\rho^2$ \\
\hline
iso-$X_7$ & $X_k,k=1,7,20,77,170,...$ & N & N & $\sum\tan(\theta_i/2)$ \\
\hline
\end{tabular}
\caption{Each of the 9 Poncelet triangle families supported by the app, and some triangle centers which remain fixed. The ``CAP'' (resp. ``CI'') column indicates if the family is interscribed in a concentric-axis parallel pair of ellipses (resp. if the outer conic is a circle). The conserved quantities in the last column are a result of experimental work, and many have been proved. Symbols $L$,$J$,$r$,$R$,$A$, $\omega$ refer to perimeter, Joachmisthal's constant \cite{sergei91}, inradius, circumradius, area, and Brocard angle, respectively. $s_i$, $\theta_i$, $\kappa_i$ refer to sidelength, internal angle, and outer ellipse curvature, respectively, $i=1,...N$, where $N$ is the number of sides of a given Poncelet family. Primed (resp. $h$-subscripteD) symbols refer to quantities measured in the excentral (resp. orthic) triangle. The quantity $\rho^2$ is the squared radius of the polar circle, negative for acute triangles \cite[Polar Circle]{mw}.}
\label{tab:n3-conc-families}
\end{table}

\subsection*{A jukebox of conjectures}

\begin{figure}
    \centering
    \includegraphics[width=.7\textwidth]{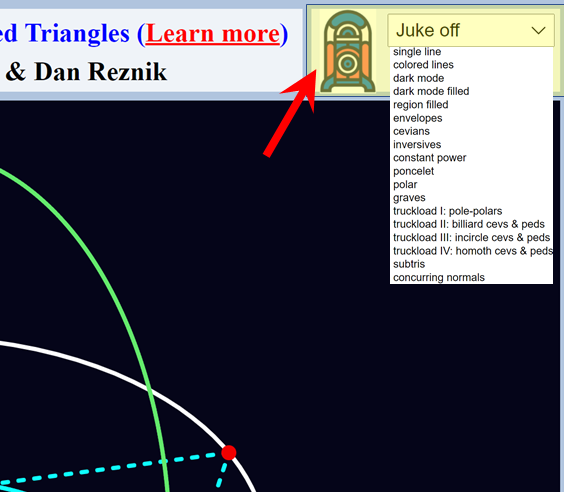}
    \caption{To play sequentially through one of many groups of experiments, select an item from the (highlighted) drop-down ``jukebox'' drop-down menu, at the top right hand corner of the app. To stop the jukebox playback, select \texttt{Juke off}. Click (or right-click) on the jukebox icon to quickly move forwards or backwards in a given sequence.}
    \label{fig:09-juke}
\end{figure}

Given the combinatorics of experiments possible, the authors have stumbled upon hundreds of curious observations (some with research value others simply of aesthetic interest). As shown in in \cref{fig:09-juke}, these are accessible as two dozen themed playlists which can be reviewed sequentially in ``jukebox mode''. Most should be regarded as experimental conjectures. All proof contributions are very much welcome!

\subsection*{Related work}

A major inspiration has been recent works investigating loci over 1d triangle families \cite{odehnal2011-poristic,pamfilos2004}.

In 2011, the second author uploaded to Youtube a video of a family of Poncelet triangles in the confocal pair \cite{reznik2011-incenter}. What was visually apparent was that the locus of the incenter was an ellipse. In the few years to follow, 3 proofs to this phenomenon were published, each with a unique approach  \cite{corentin2021-circum,garcia2023-loci,olga14}

In 2019 research resumed, now with an effort to identify conserved quantities both for $N=3$ and $N>3$ Poncelet families. Curious new observations have inspired a few recent publications \cite{akopyan2020-invariants,bialy2020-invariants,caliz2020-area-product,stachel2021-billiards}.

From our side, this has given rise to several new publications (see for example \cite{garcia2023-loci,helman2020-intriguing,reznik2020-n3-focus-inversive,reznik2021-circum,reznik2020-ballet,reznik2020-intelligencer}), and some 400 new Youtube videos of experiments; see  \cite{reznik2021-observable-media} for a complete searchable list.

Emerging themes from our experimental work have been (i) what determines if the locus of a triangle center is a conic or not, treated theoretically here \cite{helman2021-power-loci,helman2021-theory}, and (ii) can different Poncelet families be grouped according to their Euclidean invariants and/or loci \cite{garcia2020-similarity-I,reznik2020-similarityII,garcia2020-family-ties}.

\subsection*{Article organization}

In the sections below we describe the typical workflow of setting up a simulation, as well as all the combinatorics of its user-interface. The reader is also encouraged to take look at our video walk-through in \cite{reznik2021-locus-app-tutorial}.

%% file: 040_secs.tex
\section{Typical Usage Pattern}

A typical screenshot of the application is depicted in \cref{fig:09-screenshot}. A large area called here the animation window is where the dynamic geometry of a particular triangular family and its associated loci are drawn To its left is a strip of channel controls, comprising four identical groups, which define which objects are to be used as a basis to compute and draw loci from. 

The most common usage pattern is depicted in \cref{fig:09-flow}, namely: the user selects (i) a triangle family (Poncelet or ellipse-mounted, see below); (ii) the triangle on which computations will be made (the default is ``reference'' but dozens of derived triangles can be chosen); (iii) the locus type, i.e., whether one wishes to trace out a triangle center, a vertex, an envelope, etc., and (iv) which triangle center should the locus be drawn for. The first one thousand triangle centers listed in \cite{etc} are currently supported.

\begin{figure}
    \centering
    \includegraphics[width=\textwidth]{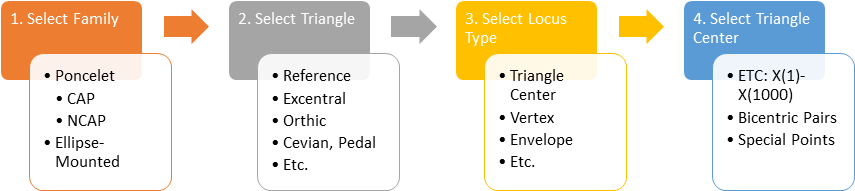}
    \caption{Caption}
    \label{fig:09-flow}
\end{figure}

In the sections below we describe the main functions of the user interface.

\section{Main ellipse and Animation Controls}
\label{sec:09-ellipse-anim}

\begin{figure}
    \centering
    \includegraphics[width=\textwidth]{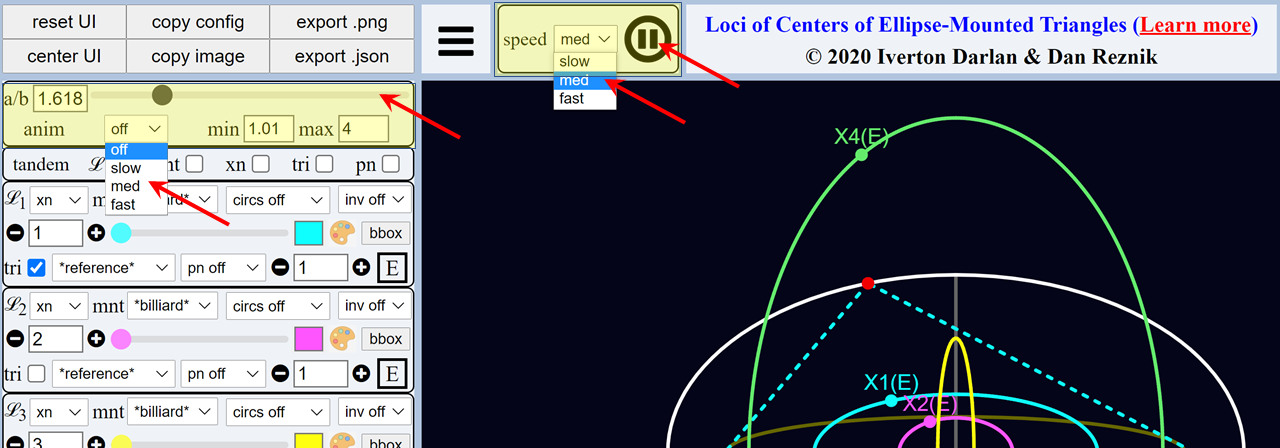}
    \caption{Basic animation controls include (i) the setting of the base ellipse aspect ratio \texttt{a/b} either via typing into the textbox (showing $1.618$ in the picture) or via the scrollbar next to it; (ii) above the animation area, pausing or running the animation and choosing a speed -- slow, medium, or fast. Note: a small ``anim'' dropbox located below the \texttt{a/b} scrollbar, when not in the ``off'' position, triggers a smooth oscillation of the aspect ratio over the range specified in the ``min'' and ``max'' input boxes to its right.}
    \label{fig:09-animation}
\end{figure}

Before a particular triangle family can be setup and its loci visualized, one must set certain basic animation controls, using the various areas highlighted in \cref{fig:09-animation}. These include (i) the setting of the base ellipse aspect ratio \texttt{a/b} either via typing into the textbox (showing $1.618$ in the picture) or via the scrollbar next to it; (ii) above the animation area, pausing or running the animation and choosing a speed -- slow, medium, or fast. Note: a small ``anim'' dropbox located below the \texttt{a/b} scrollbar, when not in the ``off'' position, triggers a smooth oscillation of the aspect ratio over the range specified in the ``min'' and ``max'' input boxes to its right.

\subsection{Convenience Animation Controls}

If the animation is paused, hitting the up (or right) and down (or left) arrows on the keyboard allows one to carefully step forward or backward over the triangle family.

The mouse wheel allows for the simulation image to be zoomed or unzoomed.

By clicking and dragging into the main animation area one can pan and reposition the image.

\subsection{Channel Controls}

\begin{figure}
    \centering
    \includegraphics[width=.6\textwidth]{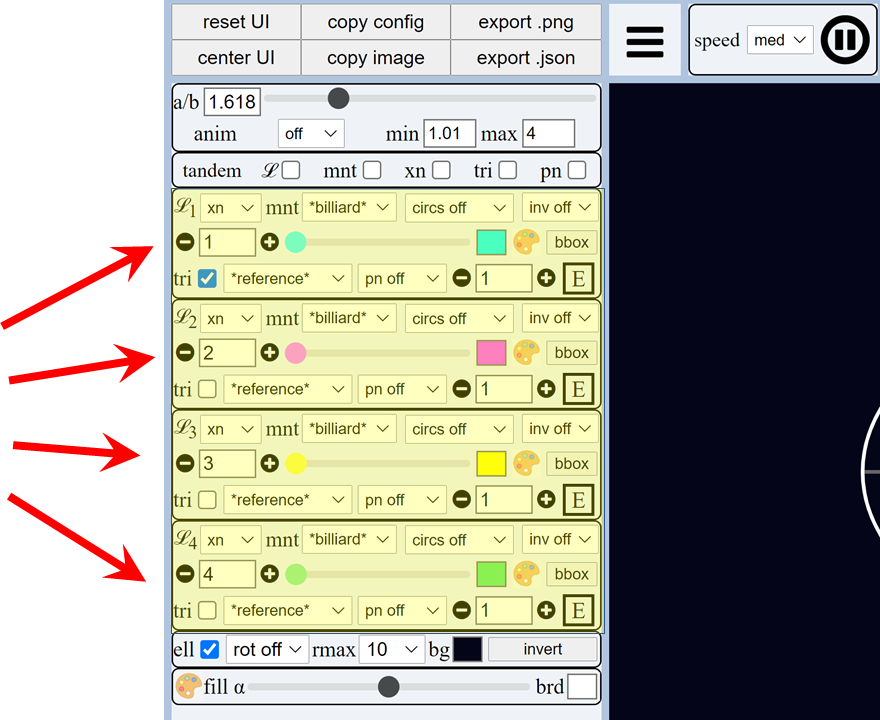}
    \caption{Four identical groups of ``channel'' controls positioned to theleft of the main animation window.}
    \label{fig:09-four-channels}
\end{figure}

\begin{figure}
    \centering
    \includegraphics[trim=0 0 150 0,clip,width=.6\textwidth]{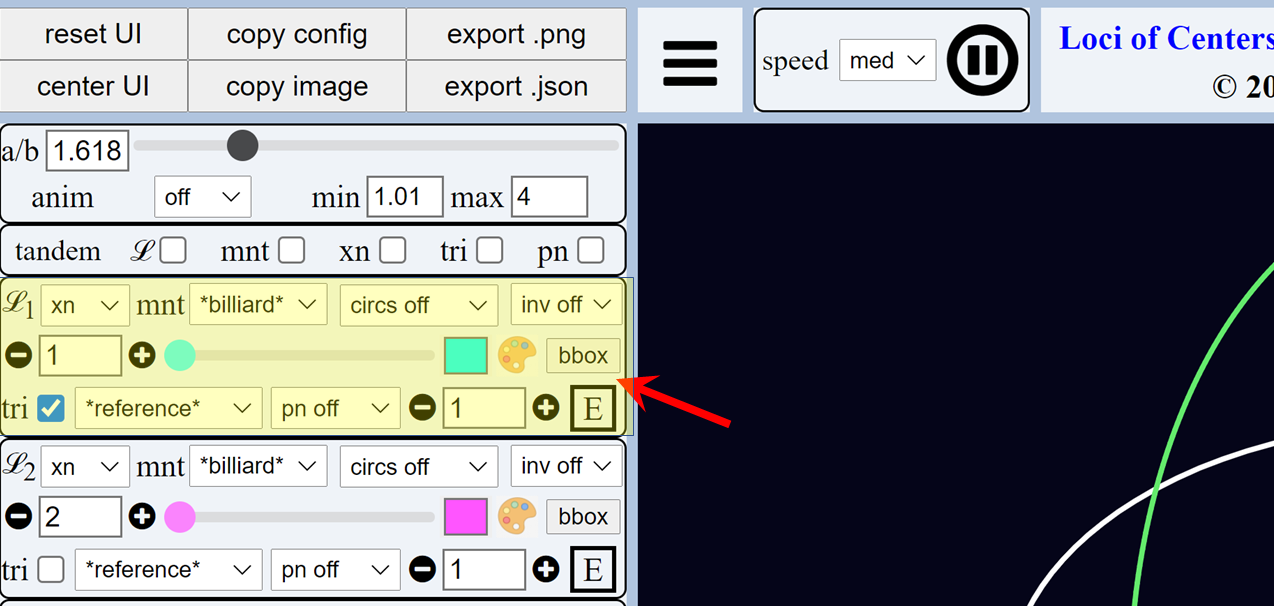}
    \caption{Various settings in a single channel control.}
    \label{fig:09-single-channel}
\end{figure}

As shown in \cref{fig:09-four-channels}, four identical groups of ``channel'' controls are positioned to the left of the main animation window. \cref{fig:09-single-channel} zooms in one of them, whose individual settings are explained next.

\section{Choosing a triangle family}

The first step in \cref{fig:09-flow} is the choice of a triangle {\em family}. A specific one is selected 
via the \texttt{mnt} drop-down, see \cref{fig:09-menu-family}. Two types of families are supported: (i) Poncelet, and (ii) ellipse ``mounted'' (see below), which originated the name of the control.

\begin{figure}
    \centering
    \includegraphics[width=.6\textwidth]{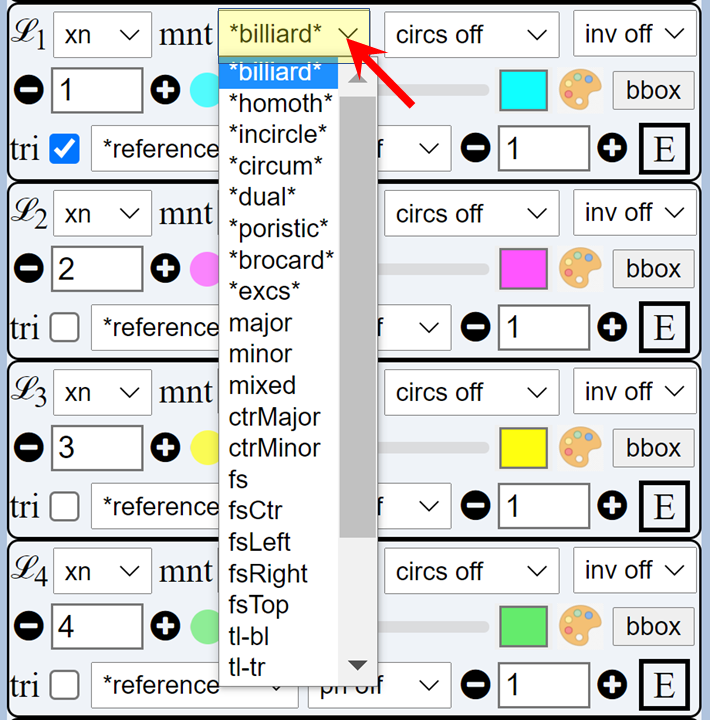}
    \caption{The \texttt{mnt} drop-down selects a triangle family.}
    \label{fig:09-menu-family}
\end{figure}

\subsection{Poncelet families} Currently we support the following types of 3-periodic Poncelet families interscribed between axis-parallel ellipses, whose names are familiar from previous sections:

\begin{itemize}
\item Ellipse-inscribed, concentric and aligned caustic, \cref{fig:six-caps}: (i) Confocal (i.e., elliptic billiard), (ii) Homothetic, (iii) with Incircle, (iv) Dual, (v) Excentral (to confocals).
\item Circle-inscribed, \cref{fig:por-broc}: (vi) Bicentric, (vii) Brocard Porism, (viii) MacBeath (excentrals to bicentric), (ix) Concentric inellipse.
\item Ellipse-inscribed, non-concentric, circular caustic, \cref{fig:circ-caustic}, where circ-$X_k$ implies that $X_k$ is stationary:
\begin{enumerate}
\item \texttt{circ}-$X_1$ (aka. `incenter-focal'), tangentials to the bicentric family;
\item \texttt{circ}-$X_2$ (aka. `iso-baric');
\item \texttt{circ}-$X_4$ (aka. `ortho-focal');
\item \texttt{circ}-$X_7$ (aka. `iso-Gergonne'), tangentials to the Brocard family.
\item Additionally, \texttt{circ-Half} and \texttt{circ-Third} families are included where the circular caustic is centered at $(a/2,b/2)$ and $(a/3,b/3)$ respectively.
\end{enumerate}

\end{itemize}

\subsection{Ellipse ``Mounted''}

Also selectable are triangle families $\Tm(t)=V_1 V_2 P(t)$, where $V_1,V_2$ are pinned to two points on or near an ellipse, and $P(t)=[a\cos{t},b\sin{t}]$ sweeps the boundary. Let The following fixed locations for $V_1$ and $V_2$ are currently supported:

\begin{enumerate}
\item \texttt{major}: left and right ellipse vertices (EVs)
\item \texttt{minor}: top and bottom EVs
\item \texttt{mixed}: left and top EVs
\item \texttt{ctrMajor}: center and left EV
\item \texttt{ctrMinor}: center and top EV 
\item \texttt{fs}: the 2 foci $f_1$ and $f_2$
\item \texttt{fsCtr}: center and right focus ($f_2$)
\item \texttt{fsLeft}: left EV and $f_2$
\item \texttt{fsRight}: right EV and $f_2$
\item \texttt{fsTop}: top EV and $f_2$
\item \texttt{tl-bl}: top left corner of ellipse bounding box (TL) and bottom left of the same (BL)
\item \texttt{tl-tr}: TL and top right corner (TR) of ellipse bounding box
\item \texttt{tl-l}: TL and left EV
\item \texttt{tl-t}: TL and top EV
\item \texttt{tl-b}: TL and bottom EV
\item \texttt{tl-o}: TL and center of ellipse
\item \texttt{tl-br}: TL and center of ellipse
\end{enumerate}

\section{Triangle Type}
\label{sec:09-triangle-type}

\begin{figure}
    \centering
    \includegraphics[width=.6\textwidth]{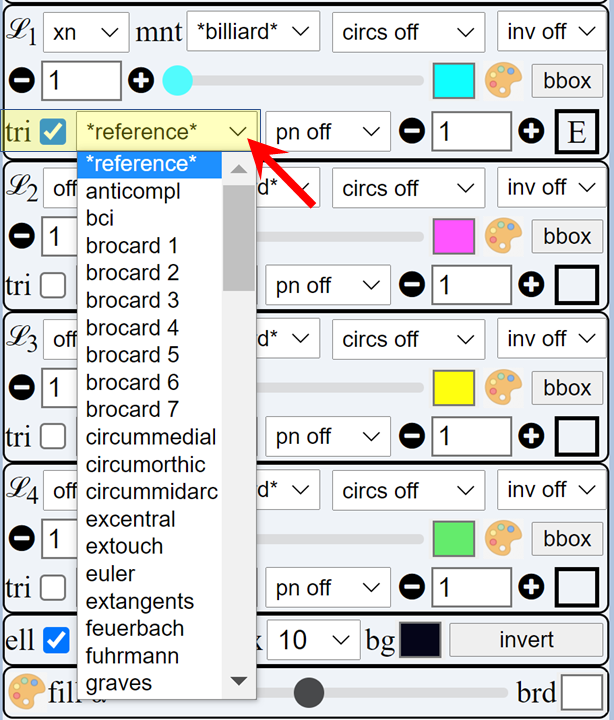}
    \caption{The triangle menu selects whether a \texttt{*reference*} or some derived triangle should be used to compute loci. The \texttt{tri} checkbox immediate to the left selects whether the triangle should be drawn or not.}
    \label{fig:09-menu-triangle}
\end{figure}

The second step in \cref{fig:09-flow} is the choice of the type of triangle with respect to which centers and loci will be computed. This is done the \texttt{tri}  checkbox and drop-down, as shown in \cref{fig:09-menu-triangle}.

\begin{figure}
\centering
\includegraphics[width=.8\textwidth]{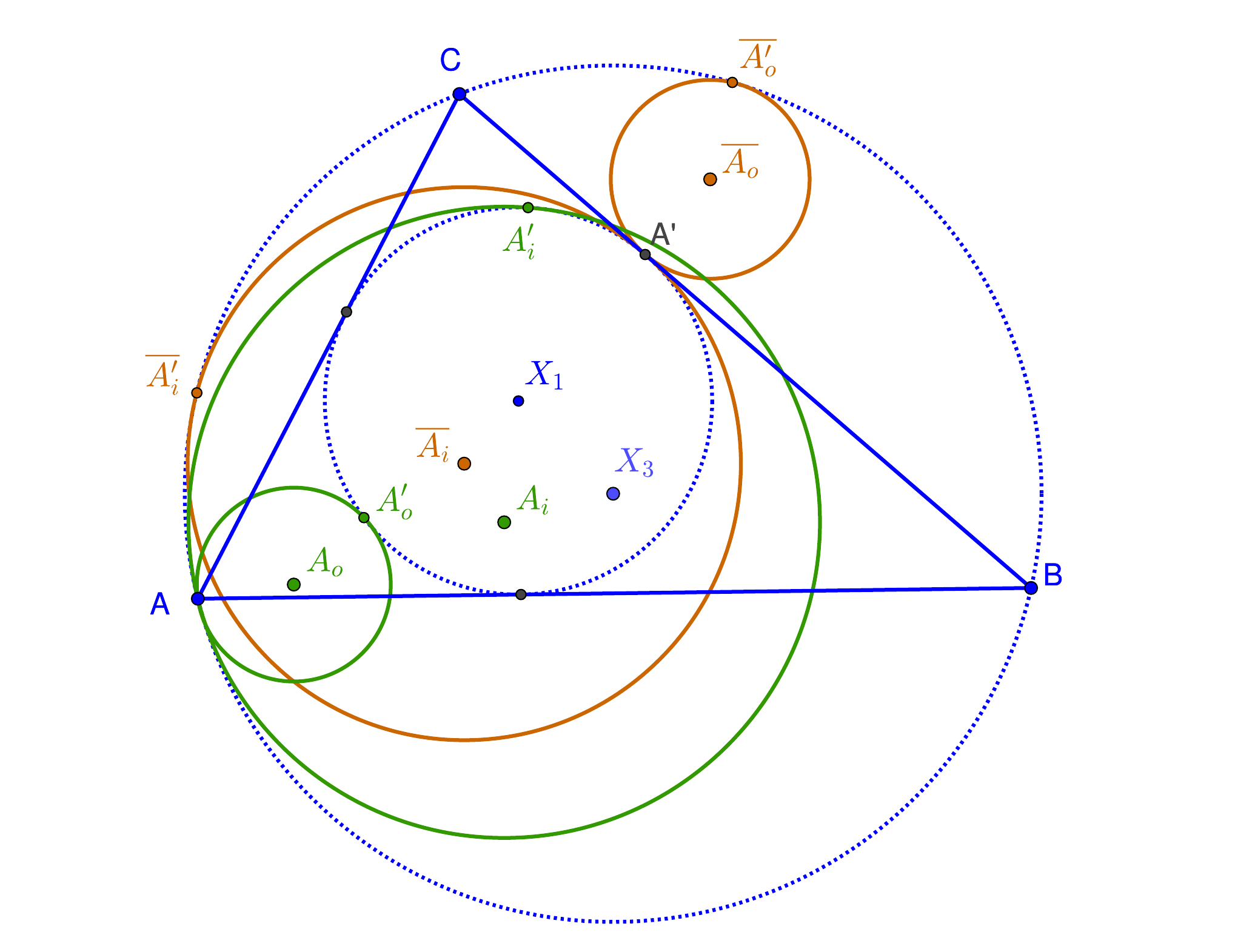}
\caption{$A$-vertices of four bi-tangent triangles derived from  $\mathcal{T}=ABC$ (blue). (i) $A_o$ and $A_i$ are centers of (green) circles bi-tangent to the circumcircle at $A$ and externally (resp. internally) to the incircle at $A_o'$ and $A_i'$, respectively. Said centers are the $A$-vertices of \texttt{.btgCext} and  \texttt{.btgCint}, respectively. (ii) $\overline{A_o}$ and $\overline{A_i}$ are the centers of (brown) circles externally (resp. internally) tangent to the incircle at the intouchpoint $A'$, while simultaneously tangent to the circumcircle (at $\overline{A_o'}$ and $\overline{A_i'}$, respectively). They are the $A$-vertices of  \texttt{.btgIext} and \texttt{.btgIint}, respectively.}
\label{fig:bitangs}
\end{figure}

While the checkbox controls whether selected triangle is drawn or not, the drop-down contains some four-dozen derived triangles. Below the default setting \texttt{*reference*} (this indicates a plain triangle in the family should be used), the choices are organized in three groups:

\begin{enumerate}
\item Standard ``named'' triangles (undecorated abbreviations), such as \texttt{anticompl} for anticomplementary, \texttt{bci} for BCI triangle, etc., whose construction can be looked up in \cite{mw}.
\item Exotic triangles (prefixed by a ``.''): \texttt{.andromeda}, \texttt{.antlia}, etc., obtained from \cite{lozada2016-triangles}.
\item Bitangent triangles\footnote{Thanks to Peter Moses for deriving their vertices}: referring to \cref{fig:bitangs}:
    \begin{itemize}
\item \texttt{.btgCext} and \texttt{.btgCint}: vertices are centers of circles internally tangent to the circumcircle at the vertices and externally (resp. internally) tangent to the incircle.
\item \texttt{.btgCextT} and \texttt{.btgCintT}: have vertices at the tangency points on the incircle used to find \texttt{.btgCext} and \texttt{.btgCint}, respectively.
\item \texttt{.btgIext} and \texttt{.btgIint}: vertices are centers of circles internally tangent to the circumcircle and externally (resp. internally) tangent to the incircle at the intouchpoints (vertices of the intouch triangle).
\item \texttt{.btgIextT} and \texttt{.btgIintT} have vertices at the tangency points on the circumcircle used to find \texttt{.btgIext} and \texttt{.btgIint}, respectively.
    \end{itemize}
\item Inversive triangles, e.g., \texttt{*inv-f1*}, \texttt{*inv-f1c*}, etc. (decorated with asterisks). 
\end{enumerate}

Below we document triangles both in the ``standard'' and ``exotic'' groups:

\subsection{Standard Triangles}

These include: Reference, Anticomplementary, BCI, 1st Brocard, 2nd Brocard, 3rd Brocard, 4th Brocard, 5th Brocard, 6th Brocard, 7th Brocard, Circum-Medial, Circum-Mid-arc, Circum-Orthic, Excentral, Extouch, 2nd Extouch\footnote{Triangle with sides bounded by the touchpoints of the excircles with the sidelines other than the extouchpoints \cite{lozada2016-triangles}.}, Extangents, Feuerbach, Fuhrmann, Half-Altitude, Hexyl, Incentral, Inner Vecten, Intangents, Intouch, Johnson, Lemoine, Lucas Central, Lucas Inner, Lucas Tangents, MacBeath, Medial, Mid-arc, Mixtilinear, 1st Morley Adj, 2nd Morley Adj, 3rd Morley Adj, 1st Neuberg, 2nd Neuberg, Orthic, Outer Vecten, Polar\footnote{With respect to the outer Poncelet conic.}, Reflection, Steiner, Symmedial, Tangential, Tangential Mid-Arc, Yff Central, Yff Contact.

\subsection{Exotic Triangles}

\begin{figure}
\centering
\includegraphics[width=.8\textwidth]{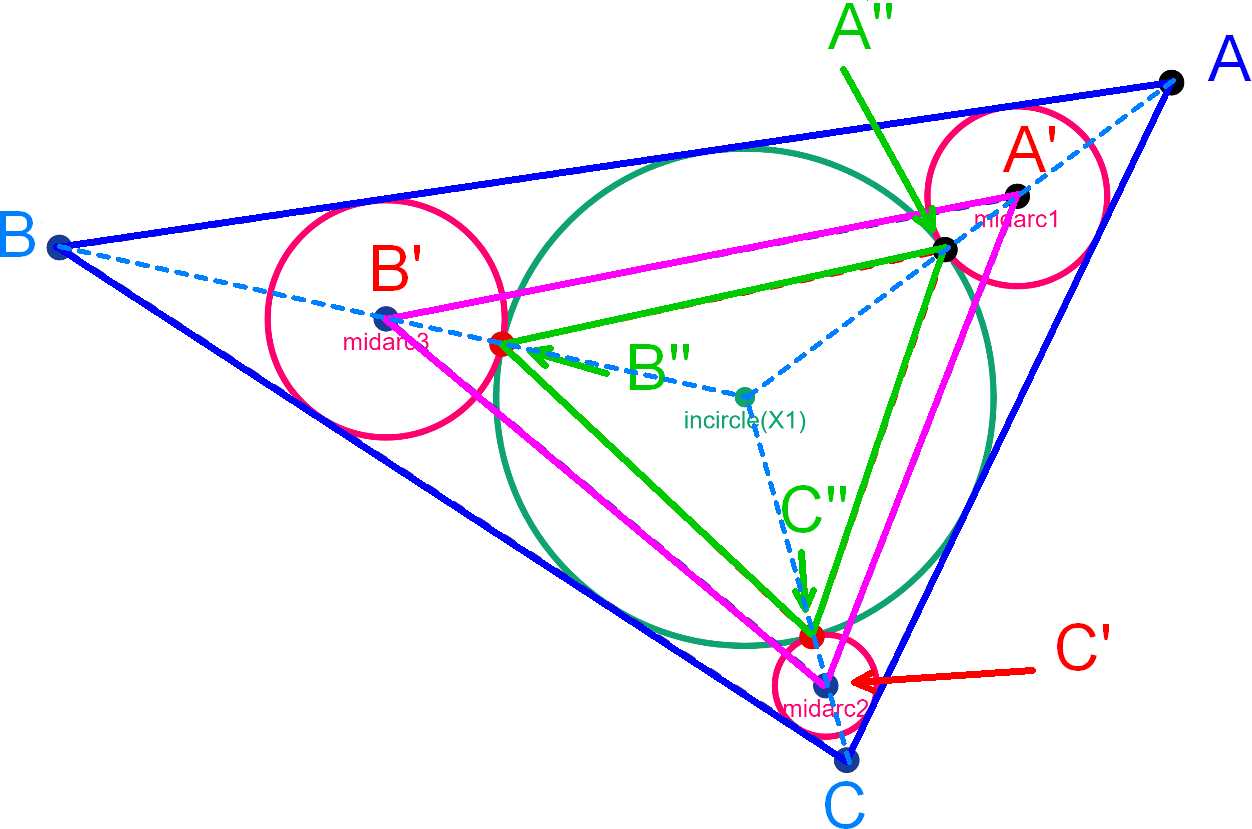}
\caption{Let $T=ABC$ be a reference triangle, shown with its incircle (green). The Mid-Arc triangle (green) $T''=A''B''C''$ has vertices at the proximal intersections of angular bisectors (dotted blue) and the incircle. The Mid-Arc circles (red) are externally-tangent to the incircle and pairs of sidelines. Their centers $A',B',C'$ are the vertices of the ``Mid-Arc circle'' triangle.}
\label{fig:mid-arc-circs}
\end{figure}

These follow definitions in \cite{lozada2016-triangles} and include: Andromeda, Antlia, Apollonius, Apus, Atik, Ayme, Bevan-Antipodal, 1st Circumperp, 2nd Circumperp, Excenters--Incenter, Reflections, Excenters--Midpoints, Honsberger, Inverse--in--Excircles, Inverse--in--Incircle, Kosnita, Mandart Excircles, Mandart Incircles, 2nd Mixtilinear, 3rd Mixtilinear, 4th Mixtilinear, Ursa Major, Ursa Minor. The following do not yet appear in \cite{lozada2016-triangles}, but are defined as follows:

\begin{itemize}
\item \texttt{8th Mixtilinear}: touchpoints of the Apollonius' circle internal to the 3 mixtilinear incircles.
\item \texttt{9th Mixtilinear}: touchpoints of the Apollonius' circle external to the 3 mixtilinear excircles.
\item \texttt{Poincaré}: has vertices at the centers of the three circles which are projections of the sidelines on the Poincaré disk\footnote{See \texttt{bit.ly/3vrHHJE}}, if the latter's boundary is identified with the circumcircle.
\item \texttt{Mid-Arc circles}: vertices are the centers of the three Mid-Arc circles, see \cref{fig:mid-arc-circs}.
\end{itemize}

\subsection{External triangle triples}

There are several creative constructions of triples of triangles based on a given reference. Currently the following are supported:

\begin{itemize}
\item \texttt{flank\{1,2,3\}}: these are one of the 3 flank triangles obtained from the sides of squares erected on each side of a reference triangle, see \cite{lamoen2001-flank}.
\item \texttt{ext out\{1,2,3\}}: these are the 3 ``external extouch'' triangles whose vertices are the point of contact of the excircles with the sidelines. 
\end{itemize}

\subsection{Inversions and other operations with respect to centers and foci of family conic pair}
\label{sec:09-adv-tri-families}

The options below are images of the reference triangle in a given family under certain operations with respect to points or circles centered on said points which are attached to the family's underlying ellipses (or caustic), e.g., center, focus, limiting points, etc.

\begin{itemize}
\item \texttt{*inv-ctr*,*inv-f1*,*inv-f2*,*inv-f1c*,*inv-f2c*}: a new triangle with vertices at the inversive images of reference vertices with respect to a unit circle centered on the outer ellipse center, left, and right foci, and inner ellipse left and right foci, respectively.
\item \texttt{*pol-ctr*,*pol-f1*,*pol-f2*,*pol-f1c*,*pol-f2c*}: a new triangle with vertices at the polar images of reference edges with respect to a unit circle centered on the outer ellipse center, left, and right foci, and inner ellipse left and right foci, respectively.
\item \texttt{*ped-ctr*,*ped-f1*,*ped-f2*,*ped-f1c*,*ped-f2c*}: the pedal triangle with the outer ellipse center, left, and right foci, and inner ellipse left and right foci, respectively.
\item \texttt{*antiped-ctr*,*antiped-f1*,*antiped-f2*,*antiped-f1c*,*antiped-f2c*}: the antipedal triangle with the outer ellipse center, left, and right foci, and inner ellipse left and right foci, respectively.
\item \texttt{*ped-lim2*}: this is specific to the confocal family. Computes the pedal triangle with respect to the non-focal limiting point of the bicentric family which is the polar image of the confocal family.
\item \texttt{*x3map-ctr*,*x3map-f1*,*x3map-f1c*}: consider a triangulation of the original triangle in 3 subtriangles, each of which contains two vertices of the original triangle and either (i) the center of the outer ellipse, (ii) its left focus, or (iii) the inner ellipse left focus, respectively. These transformations compute a new triangle with vertices at the circumcenter of each subtriangle.
\item \texttt{*x3inv-ctr*,*x3inv-f1*,*x3inv-f1c*}: these compute the inverses of the previous transform with respect to the same points.
\item \texttt{*crem-ctr*,*crem-f1*,*crem-f2*}: sends the reference vertices to their images under a quadratic Cremona transformation, which sends $(x,y)\rightarrow(1/x,1/y)$. The origin will be the center of the outer ellipse, its left focus, or its right focus, respectively. 
\item \texttt{*ellcev-ctr*, *ellcev-f1*,*ellcev-f1c*}: triangle obtained by intersecting cevians through (i) center, (ii) right focus, (iii) right outer focus of family's ellipse(s) with the outer ellipse. 
\item \texttt{*inf-x*,*inf-y*,*inf-x2*,*inf-y2*} (experimental): They dynamically set the $x$ or $y$ coordinate of each vertex so they slide along infinity-like Lissajous curves.
\end{itemize}
 
\section{Locus Type}

\begin{figure}
    \centering
    \includegraphics[width=.6\textwidth]{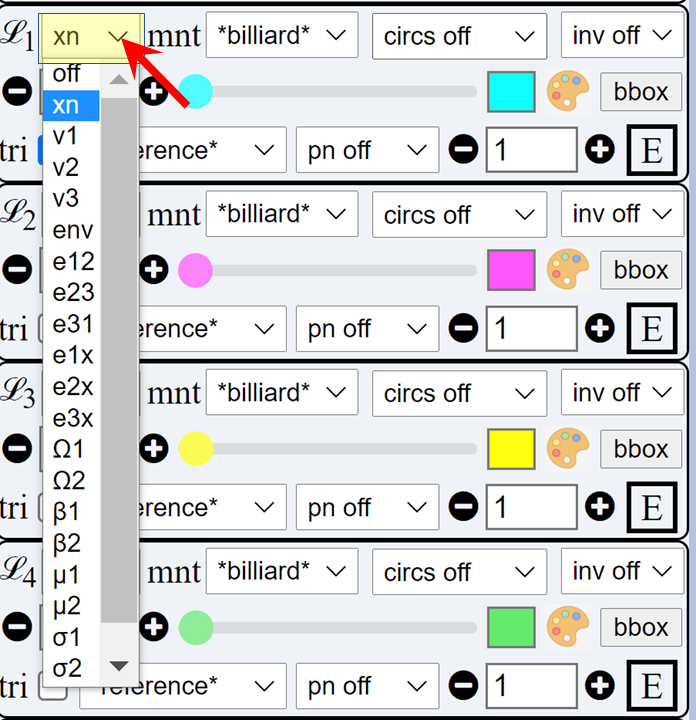}
    \caption{The $\mathcal{L}_i$ menu selects the locus type to (triangle center, vertex, envelope, etc.).}
    \label{fig:09-menu-locus}
\end{figure}

The third step in \cref{fig:09-flow} is the choice of type of locus to be drawn, or more precisely, the feature selected from the family/triangle combination previously selected. This is done with the $\mathcal{L}_i$ menu at the top of the control group, $i=1,2,3,4$, shown in \cref{fig:09-menu-locus}.

There are three conceptual groups of locus types: (i) triangle centers and vertices, (ii) segment envelopes, and (iii) bicentric pairs. These are explained next.

\subsection{Centers and Vertices}

\begin{enumerate}
\item \texttt{off}: it indicates the trace (locus) of this channel should not be drawn. It is the default setting for channels $2,3,4$ upon startup.
\item \texttt{xn}: draw the locus of the selected triangle center, as in  \cref{sec:09-triangle-center};
\item \texttt{v1}, \texttt{v2}, \texttt{v3}: show the trace of one of thee vertices of the triangle family. In Poncelet families, these will sweep out the same curve, but this is not the case for ellipse-mounted families.
\item \texttt{ort}: the {\em orthopole} of line $X_m X_n$, see \cite[Orthopole]{mw}, where $m$ and $n$ are selected triangle and cevian centers, see \cref{sec:09-triangle-center} and \cref{sec:09-cevian}.
\end{enumerate}

\subsection{Envelopes}

\begin{enumerate}
\item \texttt{env}: the envelope of segment $X_m X_n$, $m{\neq}n$, where $m$ (resp. $n$) is the selected triangle (resp. cevian) center. 
\item \texttt{e12}, \texttt{e23}, \texttt{e31}: the envelope of side $V_i V_j$ of the triangle family. Note these are one and the same (resp. distinct) for Poncelet (ellipse-mounted) families. 
\item \texttt{e1x}, \texttt{e2x}, \texttt{e3x}: the envelope of $V_i X_n$, i.e., the line from a given vertex to a selected triangle center. In a concentric Poncelet family, the envelope of $V_i X_1$ will be the outer ellipse's evolute, see it \href{https://bit.ly/3fNKV2P}{Live}.
\end{enumerate}

\subsection{Bicentric Pairs}

Only a few have so far been implemented, from the copious list in \cite{kimberling2020-bicentric}.

\begin{enumerate}
\item $\Omega_1,\Omega_2$: the Brocard points
\item $\beta_1,\beta_2$: the Beltrami points: inversions of the Brocard points with respect to the circumcircle
\item $\mu_1,\mu_2$: also known as ``Moses'' points: inversion of the Brocard points with respect to the incircle.
\item $\sigma_1,\sigma_2$: the two foci of the Steiner circumellipse (aka. the Bickart points)
\end{enumerate}

\section{Triangle Center}
\label{sec:09-triangle-center}

\begin{figure}
    \centering
    \includegraphics[width=.6\textwidth]{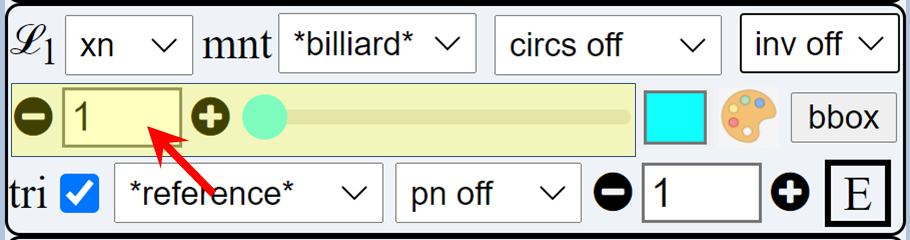}
    \caption{Controls used for the selection of a particular triangle center $X_k$.}
    \label{fig:09-menu-xn}
\end{figure}

The fourth and final step in \cref{fig:09-flow} is the choice of triangle center $X_k$ in the region highlighted in \cref{fig:09-menu-xn}. There are three ways to choose $k\in[1,1000]$: (i) by typing/editing the text field showing $k$, (ii) incrementing or decrementing $k$ by clicking on the ``-'' and ``+'' symbols around the text field; (iii) using the scrollbar to the right of the ``+'' control, to quickly scroll through all 1000 values of $k$. In fact after any of these is performed, this set of controls becomes ``focused'' in such a way that (iv) left (resp. right) arrow keystrokes will decrement (resp. increment) the value, allowing mouse-free traversal of triangle centers.

\section{Cevians, Pedals, \& Co.}
\label{sec:09-cevian}

\begin{figure}
    \centering
    \includegraphics[width=.6\textwidth]{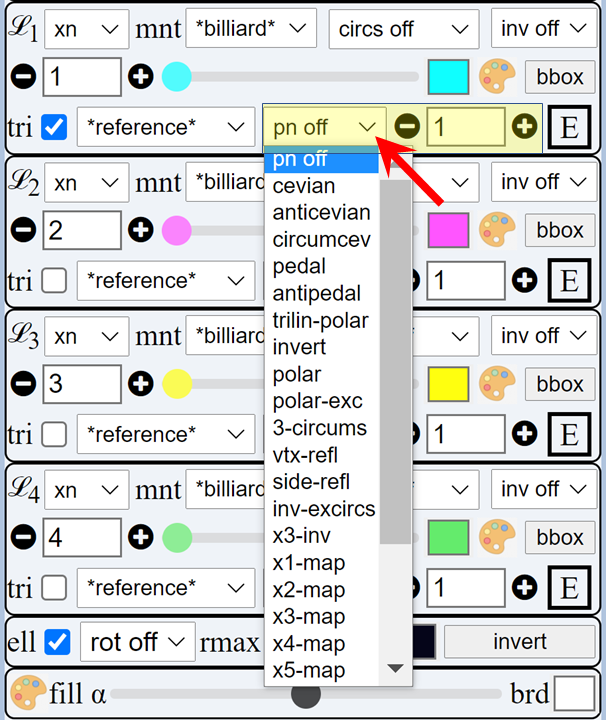}
    \caption{Cevian-like triangles and number box to select a triangle center playing the role of $Q$ (see text).}
    \label{fig:09-menu-cevian}
\end{figure}

An additional ``cevian-like'' transformation with respect to an additional triangle center $X_m$ can be applied to the triangle type selected in \cref{sec:09-triangle-type}. Let us call the latter the ``parent'' triangle.
The specific transformation is selected via the drop-down menu in \cref{fig:09-menu-cevian} (the default setting is \texttt{pn off}, meaning this additional transformation is inactive), and $X_m$ via the numeric input box to the right of the menu.

The $X_m$-transformations possible are grouped into (i) traditional, (ii) inversive, (iv) reflexive, and (iv) triangulated. Below, let $T_m$ denote the transformed triangle, and $P_i$, $i=1,2,3$, the vertices of the parent triangle.

\subsection{Traditional}

Available in this groups are the standard constructions for (i) Cevian,  (ii) Anticevian, (iii) Circumcevian, (iv) Pedal, (v) Antipedal, and (vi) Trilinear Polar triangles described in \cite{mw}. Recall that the latter produces a degenerate (segment-like) triangle, see \cite[Trilinear Polar]{mw}. Additionally, the setting \texttt{ccev-simson} computes the triangle bounded by the Simson lines corresponding to the vertices of the $X_m$-circumcevian triangle.

\subsection{Inversive}

\begin{itemize}
\item \texttt{invert}: $T_m$ will have vertices at inversions of the parent one with respect to a unit circle centered on $X_m$.
\item \texttt{polar}: $T_m$ will be bounded by the polars (infinite lines) of the parent's vertices with respect to a unit circle centered on $X_m$, see \cite[Polar]{mw}.
\item \texttt{inv-excircs}: $T_m$ will have vertices at inversions of $X_m$ with respect to its excircles, see \cite[Excircle]{mw}. 
\item \texttt{polar-exc}: $T_m$ will be bounded by the polars (infinite lines) of $X_m$ with respect to each of the parent's excircles. 
\end{itemize}

\subsection{Reflexive}

\begin{itemize}
\item \texttt{vtx-refl}: $T_m$ has vertices at the reflections of $X_m$ on the parent vertices. 
\item \texttt{side-refl}: $T_m$ has vertices at the reflections of $X_m$ on the sidelines of the parent triangle.
\end{itemize}

\subsection{Triangulated}

Triangulate the parent with respect to $X_m$, i.e., consider the following subtriangles: $T_{23} = X_m P_2 P_3$, $T_{31} = X_m P_3 P_1$, and $T_{12} = X_m P_1 P_2$.

\begin{itemize}
\item \texttt{3-circums}: $T_m$ has vertices at the circumcenters of $T_{23}$, $T_{31}$, and $T_{12}$. Note: if $X_m=X_4$, you obtain the Johnson triangle.
\item \texttt{3-inv}: The inverse of \texttt{3-circums}. $T_m$ is such that the circumcenters of its three subtriangles are the vertices of the parent. The vertices of $T_m$ are the non-$X_m$ intersections of a circle through $X_m$ and $P_i$ with a circle through $X_m$ and $P_{i+1}$, cyclically. 
\item $X_k$-map, $k\in[1,11]$: $T_m$ has vertices at the $X_k$ of $T_{23}$, $T_{31}$, and $T_{12}$. Note: $X_3$-map is the same as the \texttt{3-circums} setting.
\end{itemize}

\subsection{Pedal and Cevian Subtriangles}

\begin{itemize}
\item \texttt{subcevian,subpedal}: a triangle with one vertex from the original and two closest vertices in the $X_m$-cevian ($X_m$-pedal).
\item \texttt{subanticev,subantiped}: a triangle with one vertex from the $X_m$-anticevian ($X_m$-antipedal) and the two closest ones in the original.
\end{itemize}

\subsection{Ellipse cevians}

\begin{itemize}
\item \texttt{ell-cev}: computes the intersections of cevians thru $X_m$ with the ellipse to which the Poncelet family is inscribed. 
\item \texttt{cau-cev}: computes the intersections of cevians thru $X_m$ with the inner ellipse (caustic) of the Poncelet family. 
\end{itemize}

\subsection{Triangles from Triples}

Consider a triple of triangles derived from a reference one. A new triangle can be constructed with vertices at the $X_m$ of each in the triple. Currently we support:

\begin{itemize}
\item \texttt{flank}: $X_m$ of the three flank triangles, with vertices at squares erected above each side, see \cite{lamoen2001-flank}.
\item \texttt{extouch}: $X_m$ of the 3 ``outer extouch'' triangles, whose vertices are the touchpoints of the excircles with the sidelines, including the extouchpoints themselves.
\item \texttt{xk-subcevian,xk-subanticev}, where $k$ can be 1,2,6,7,8,9,10,11 (all interior to a triangle). These construct a triple of triangles which depart from (and exclude) the $X_m$-cevian (resp. $X_m$-anticevian).
\end{itemize}

\section{Notable Circles}
\label{sec:09-inversion-circles}

\begin{figure}
    \centering
    \includegraphics[width=.6\textwidth]{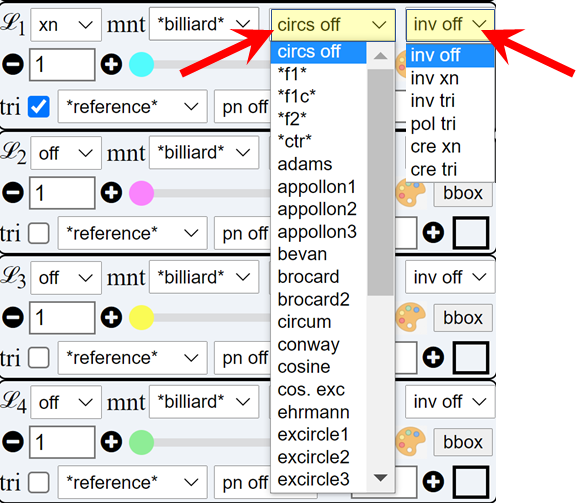}
    \caption{The left drop-down selects an ellipse-based circle or a ``central'' circle for both visualization and/or as references for inversive transformations.}
    \label{fig:09-menu-circles}
\end{figure}

Dozens of circles can be visualized with respect to the triangle family selected in \cref{sec:09-triangle-type}. These are selected via the (left) drop-down menu highlighted in \cref{fig:09-menu-circles}. The \texttt{circs off} setting is the default. The possible choices are organized in two groups: (i) ellipse-affixed, and (ii) central circles.

\subsection{Ellipse-Affixed Circles}

These are asterisk-decorated to indicate that they refer to a unit circle centered on a notable point of the ellipse (or caustic) used to generate a given triangle family, to be sure:

\begin{itemize}
\item \texttt{*ctr*}: in a pair of concentric ellipses, the common center, else the center of the circumcircle (bicentric and Brocard).
\item \texttt{*f1*,*f2*}: the left (resp. right) focus of the outer ellipse. Note: in the bicentric (resp. excentral) family this becomes the center of the outer circle (resp. caustic = elliptic billiard).
\item \texttt{*f1c*,*f2c*}: the left (resp. right) focus of the inner ellipse. 

\end{itemize}

\subsection{Central Circles}

Most of these are defined in \cite[Central Circles]{mw}:

\begin{itemize}
\item \texttt{adams}: the Adams circle
\item \texttt{apollon out}: the outer apollonius' circle (tangent and external to excircles)
\item \texttt{bevan}: the Bevan circle, circumcircle of the excentral triangle
\item \texttt{brocard,brocard2}: the Brocard circle and the so-called ``2nd'' Brocard circle.
\item \texttt{circum}: the circumcircle
\item \texttt{conway}: Conway's circle
\item \texttt{cosine}: the cosine (or 2nd Lemoine) circle 
\item \texttt{cos.exc}: the cosine circle of the excentral triangle
\item \texttt{ehrmann}: Ehrmann's 3rd Lemoine circle, see \cite{darij2012-ehrmann}.
\item \texttt{euler}: Euler's circle
\item \texttt{furhmann}: Furhmann's circle
\item \texttt{gallatly}: Gallatly's circle
\item \texttt{gheorghe}: Gheorghe's circle, see \cite[X(649)]{etc}
\item \texttt{incircle}: Incircle
\item \texttt{lemoine}: 1st Lemoine circle
\item \texttt{lester}: Lester's circle
\item \texttt{mandart}: Mandart's circle
\item \texttt{mixt.inner}: the inner Apollonius' circle to the mixtilinear incircles, i.e., it is contained and tangent to all of them
\item \texttt{mixt.exc outer} the outer Apollonius' circle to the mixtilinear excircles, i.e., it contains and is tangent to all of them
\item \texttt{moses,moses rad}: Moses's circle and Moses' radical circle
\item \texttt{parry}: Parry's circle
\item \texttt{polar}: Polar circle, centered on $X_4$. Since radius for acute triangles is negative, we show it with absolute radius) \cite[Polar circle]{mw}.
\item \texttt{reflection}: the ``reflection'' circle (circumcircle of the reflection triangle)
\item \texttt{schoutte}: Schoutte's circle
\item \texttt{spieker}: Spieker's circle
\item \texttt{taylor}: Taylor's circle
\end{itemize}

\begin{figure}
\centering
\includegraphics[width=.8\textwidth]{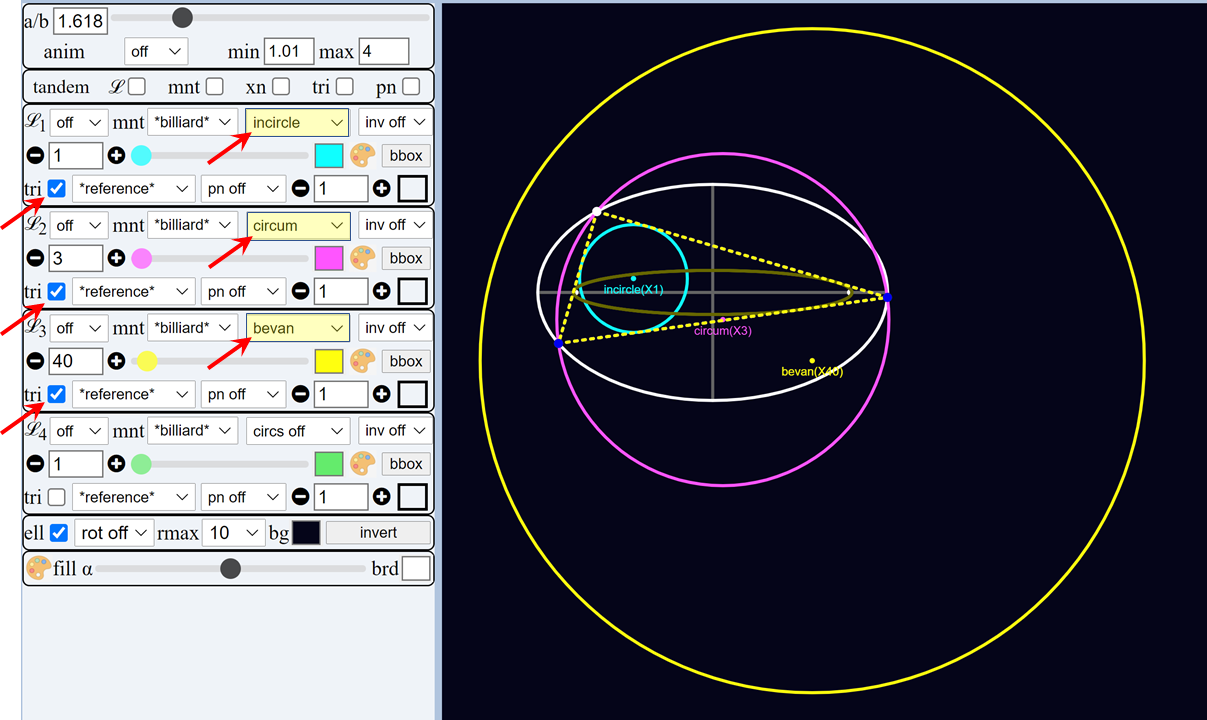}
\caption{The incircle, circumcircle, and Bevan circle are viewer simultaneously, by choosing them on the circle menu in 3 separate channels. Notice that to make the circle appear, one must check the \texttt{tri} checkbox in the lower left of that channel control area. \href{https://bit.ly/3poF5YQ}{Live}}
\label{fig:09-three-circles}
\end{figure}

As shown in \cref{fig:09-three-circles}, several circles can be shown simultaneously. To do this select one for each channel (maintaining the same triangle family and type), and make sure to check the \texttt{tri} checkboxes for each channel.

\subsection{Circle Triples}

To save on channels used to display individual circles, certain triples of circles can be shown simultaneously. These are also available in the drop-down used to select circles and include:

\begin{itemize}
\item \texttt{apolls123}: the three Apollonius' circles (passing through a vertex and the two isodynamic points)
\item \texttt{btgCext123, btgCint123}: circles bitangent to the circumcircle at the vertices and externally (resp. internally) to the incircle. Their centers are the vertices of the two bitangent triangles \texttt{.btgCext} and \texttt{.btgCint}, see \cref{sec:09-triangle-type}.
\item \texttt{btgIext123, btgIint123}: circles bitangent to the circumcircle and externally (resp. internally) to the incircle at the intouchpoints. Their centers are the vertices of the two bitangent triangles \texttt{.btgIext} and \texttt{.btgIint}, see \cref{sec:09-triangle-type}.
\item \texttt{brocards123}: the 1st Brocard, 2nd Brocard, and Moses circle
\item \texttt{excircs123}: the three excircles
\item \texttt{inc,cir,eul}: the incircle, circumcircle, and Euler circle
\item \texttt{johnsons123}: the three Johnson circles
\item \texttt{lemoines123}: the 1st Lemoine, 2nd Lemoine (aka. cosine circle), and 3rd Lemoine (aka. Ehrmann circle \cite{darij2012-ehrmann})
\item \texttt{midarcs123}: the Mid-Arc circles, tangent to pairs of sides and the incircle, \cref{fig:mid-arc-circs}.
\item \texttt{mixts123} (resp. \texttt{mixt excs123}: the three mixtilinear incircles (resp. excircles) 
\item \texttt{neubs.123} (resp.  \texttt{neubs.123r}): the three Neuberg (resp. reflected Neuberg) circles
\item \texttt{poincs123}: the three circles which are projections of the sidelines on the Poincaré disk\footnote{See \texttt{bit.ly/3vrHHJE}}, if the latter's boundary is identified with the circumcircle. 
\item \texttt{powers123}: the three ``power circles'' (centered on opposing sides' midpoints and passing through a given vertex)
\end{itemize}

\subsection{First of triples}

To only show the $A$-vertex specimens of each of the above triples use:

\begin{itemize}
\item \texttt{appollon1:} first Apollonius' circle (contains one vertex and the isodynamic points)
\item \texttt{excircle1}: excircle opposite to a first vertex
\item \texttt{johnson1}: 1st Johnson circle
\item \texttt{mid-arcs1}: 1st Mid-Arc circle, \cref{fig:mid-arc-circs}.
\item \texttt{mixtilin1}: mixtilinear incircle tangent to two sides and the circumcircle, see \cite{yiu2005-elegant,rabinowitz2006-mixt}
\item \texttt{mixtilin exc1}: first mixtilinear excircle, see \cite{nguyen2006-mixt-exc}
\item \texttt{neuberg1,neuberg1r}: 1st Neuberg circle and its side-reflected image
\item \texttt{power1}: 1st power circle (center on a side's midpoint and through two opposing vertices)
\end{itemize}

\section{Inversive Transformations with respect to a Circle}

Provided a single circle $\Cm$ (or a triple thereof) is selected as in the above, one can add an inversive-type transformation with respect to it. This is done via the (right) drop-down menu highlighted in \cref{fig:09-menu-circles}. The possible transformations are as follows: 

\begin{itemize}
\item \texttt{inv off}: no inversion is performed.
\item \texttt{inv ref}: no inversion, but circle shown will be always wrt to reference triangle (and not a derived one if one selected from the triangle type menu).
\item \texttt{inv xn}: invert the selected triangle center (see \cref{sec:09-triangle-center}) with respect to $\Cm$.
\item \texttt{inv tri}: invert the vertices of triangles in the family with respect to $\Cm$.
\item \texttt{pol tri}: compute a new triangle bounded by the polars of the original vertices with respect to $\Cm$.
\item \texttt{pol.sd}: compute a new triangle with vertices at the poles of original sides with respect to $\Cm$.
\item \texttt{cre xn}: send the selected triangle center to its image under a quadratic Cremona transformation (QCT) $(x,y)\rightarrow(1/x,1/y)$, where $(x,y)$ are the coordinated of the center of $\Cm$. 
\item \texttt{cre tri}: compute a new triangle whose vertices are images of the QCT with respect to the center of $\Cm$.
\end{itemize}

\section{Conic and Invariant Detection}

\subsection{Curve Type}

\begin{figure}
    \centering
    \includegraphics[width=.8\textwidth]{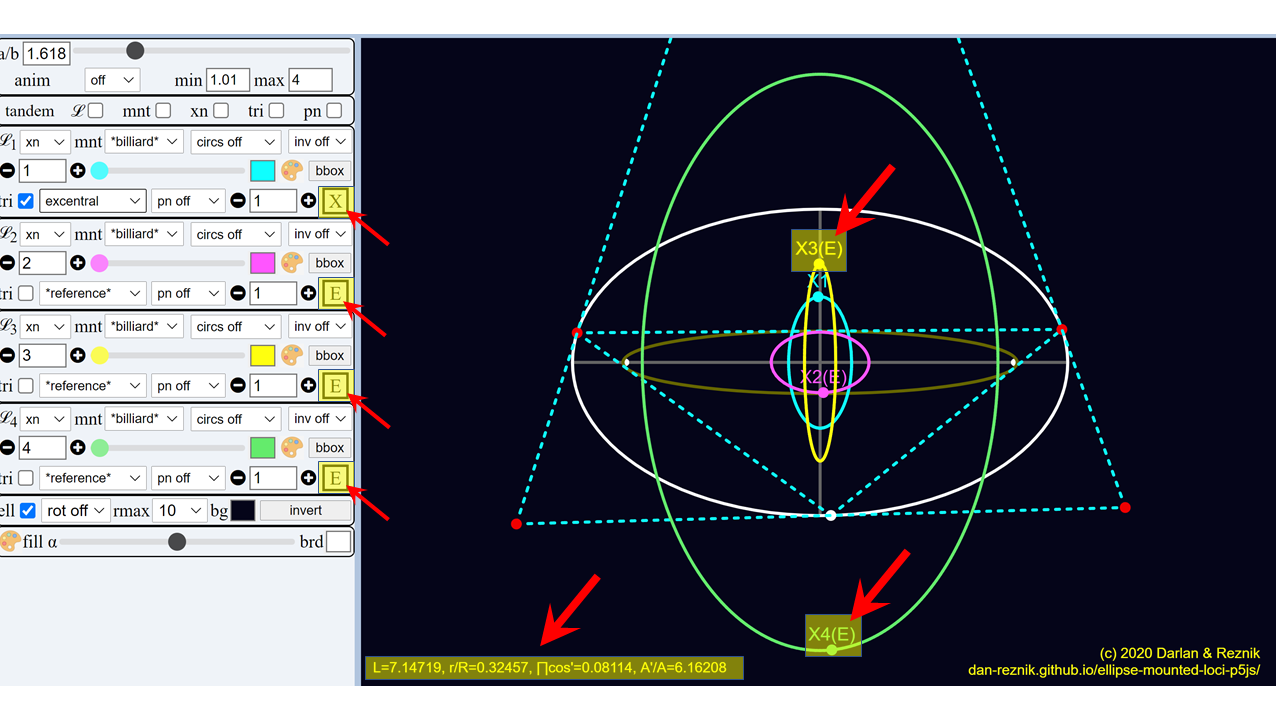}
	\caption{An indication as to curve type of each locus appears in a small box in the lower right-hand side of each control group. In the picture, \texttt{X} means the first locus (incenter over the excentral family) is non-conic. An \texttt{E} in the remainder 3 channels indicates their loci are ellipses. Notice the same indicator is appended to the instantaneous location of the triangle centers being tracked, e.g., \texttt{X3(E)} indicates the locus of the circumcenter is an ellipse. \href{https://bit.ly/3uXndpa}{Live}}
    \label{fig:09-conic-invariants}
\end{figure}

As shown in \cref{fig:09-conic-invariants}, when one or more loci are displayed, the app  indicates in the lower right-hand side of the corresponding control group, the curve type of the locus (detected via least-squares curve fitting). The following codes are used:

\begin{itemize}
\item \texttt{X}: non-conic
\item \texttt{E}: ellipse
\item \texttt{H}: hyperbola
\item \texttt{P}: parabola (very rare)
\item \texttt{L}: line or segment
\item \texttt{*}: a stationary point.
\end{itemize}

The same code is also appended (in parenthesis) to the (moving) triangle center being displayed, for example, \cref{fig:09-conic-invariants}, \texttt{X2(E),X3(E),X4(E)}, indicate the loci of barycenter, circumcenter, and orthocenter are ellipses over billiard 3-periodics.

\subsection{Detection of Metric Invariants}

The app also reports when certain basic, metric quantities are invariant, currently over triangles in the first channel only. These appear as a single line at the bottom of the animation area of a given experiment, see \cref{fig:09-conic-invariants}. In the example, the following line of text is reported:

\[ L=7.14..., r/R=0.32..., \prod\cos'=0.0811, A'/A=6.6... \]

In turn, this means that perimeter $L$ and ratio $r/R$ of inradius-to-circumradius are  numerically invariant over the reference family selected in channel 1, and that the product $\prod\cos'$ of cosines, and ratio $A'/A$ of derived-by-reference areas is constant (these are observations first introduced in \cite{reznik2020-intelligencer}).

Reported invariants appear unprimed to refer to the reference triangle in a given family. Primed quantities will appear when a derived triangle has been selected (e.g., ``excentral''), allowing for mutual comparison. 

The following quantities are currently reported, when numerically invariant:

\begin{itemize}
\item $L,A$: perimeter (sum of sidelengths $\sum{s_i}$) and area
\item $\sum{s_i^{-2}},\sum{s^{2}_i},\sum{s_i^{-2}},\sum{\sqrt{s_i}},\sum{1/\sqrt{s_i}}$: sum of inverse, squared, inverse-squared, square-root, and inverse-square-root sidelengths, respectively.
\item $\sum{s_{ij}},\sum{s_{ij}^{-2}},\sum{s^{2}_{ij}},\sum{s_{ij}^{-2}}$: sum of plain, inverse, squared, and inverse-squared `cyclic' sidelength products, $\{s_{ij}\}=\{s_1 s_2, s_2 s_3, s_3 s_1\}$, respectively.
\item $\prod{s_i}$: product of sidelengths.
\item $r,R, r/R$: inradius, circumradius, and their ratio. Latter is equivalent to invariant sum of cosines since $\sum{\cos_i}=1+r/R$.
\item $\cot\omega$: the cotangent of the Brocard angle (equivalent to constant sum of cotangents since $\sum{\cot_i}=\cot{\omega}$).
\item $\sum{\sin},\sum{\cos},\sum{\tan}, \sum\sec,\sum\csc$: sum of internal angle sines, cosines, tangents, secants, and co-secants.
\item $\sum{\sin(2t)},\sum{\cos(2t)},\sum{\tan(2t)},\sum{\sec(2t)},\sum{\csc(2t)},\sum{\cot(2t)}$: sum of double-angle trigonometric quantities
\item $\sum{\sin(t/2)},\sum{\cos(t/2)},\sum{\tan(t/2)},\sum{\sec(t/2)},\sum{\csc(t/2)},\sum{\cot(t/2)}$: sum of half-angle trigonometric quantities.
\item $\prod\sin,\prod\cos\prod\cot$: the product of internal angle sines, cosines, and cotangents.
\item $R_c$: reports if the radius of a circle selected via the circle menu (\cref{sec:09-inversion-circles}) is constant.
\item $r_{pol}^2$: squared-radius of the triangle's ``polar circle'', $r_{pol}^2=4R^2-(\sum{s_i}^2)/2$, negative for acute triangles \cite[Polar Circle]{mw}.
\end{itemize}

Also reported:

\begin{itemize}
\item When a derived triangle is selected: invariant $L'/L$, $A'/A$, $A'.A$, $R/R'$ ratios.
\item When a central circle is selected: invariant radius $R_c$.
\item When both a derived triangle and central circle are selected: $R_c'/R_c$ is constant.
\end{itemize}

\section{The tandem bar}

A common exploratory pattern is to observe the behavior of loci across all channels simultaneously while a single setting is varied, e.g., triangle family, triangle type, etc. This could be done with tedious mouse-based changes (of the varying parameter) across all controls. 

As an example, consider observing the loci of $X_k$, $k=1,2,3,4$ for the billiard family and then for the homothetic family. This would require one to reset each of the four \texttt{mnt} drop-downs from \texttt{billiard} to \texttt{homothetic}. If the user now wished to examine said loci over the incircle family, all \texttt{mnt} drop-downs wouldhave to be reset to \texttt{incircle}, etc. 

Referring to \cref{fig:09-tandem-bar}, the tandem bar, makes this rather common usage pattern very efficient.Namely, the user can set one or more tandem checkboxes causing a given setting to be ``short circuited'' across all channels. Specifically: 

\begin{figure}
    \centering
    \includegraphics[width=.8\textwidth]{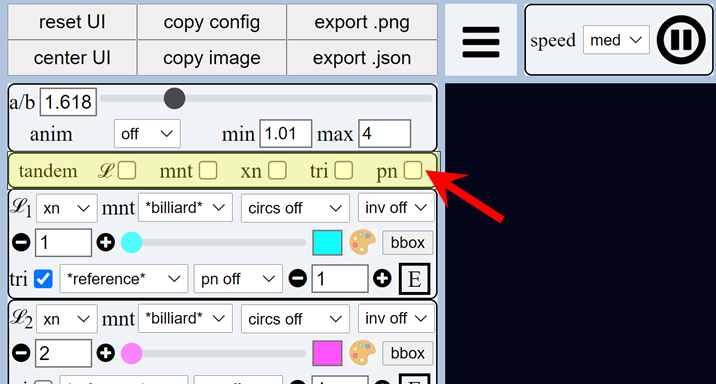}
    \caption{The tandem bar. Checking one or more checkboxes ensures locks all corresponding drop-downs in the channel controls to take the same value.}
    \label{fig:09-tandem-bar}
\end{figure}

\begin{itemize}
\item $\Lm$: the locus type
\item \texttt{mnt}: the Poncelet or ellipse-mounted family
\item \texttt{xn}: the triangle center number
\item \texttt{tri}: the reference or derived triangle
\item \texttt{pn}: the triangle center with respect to which cevian-like triangles are calculated
\end{itemize}

\begin{figure}
    \centering
    \includegraphics[width=\textwidth]{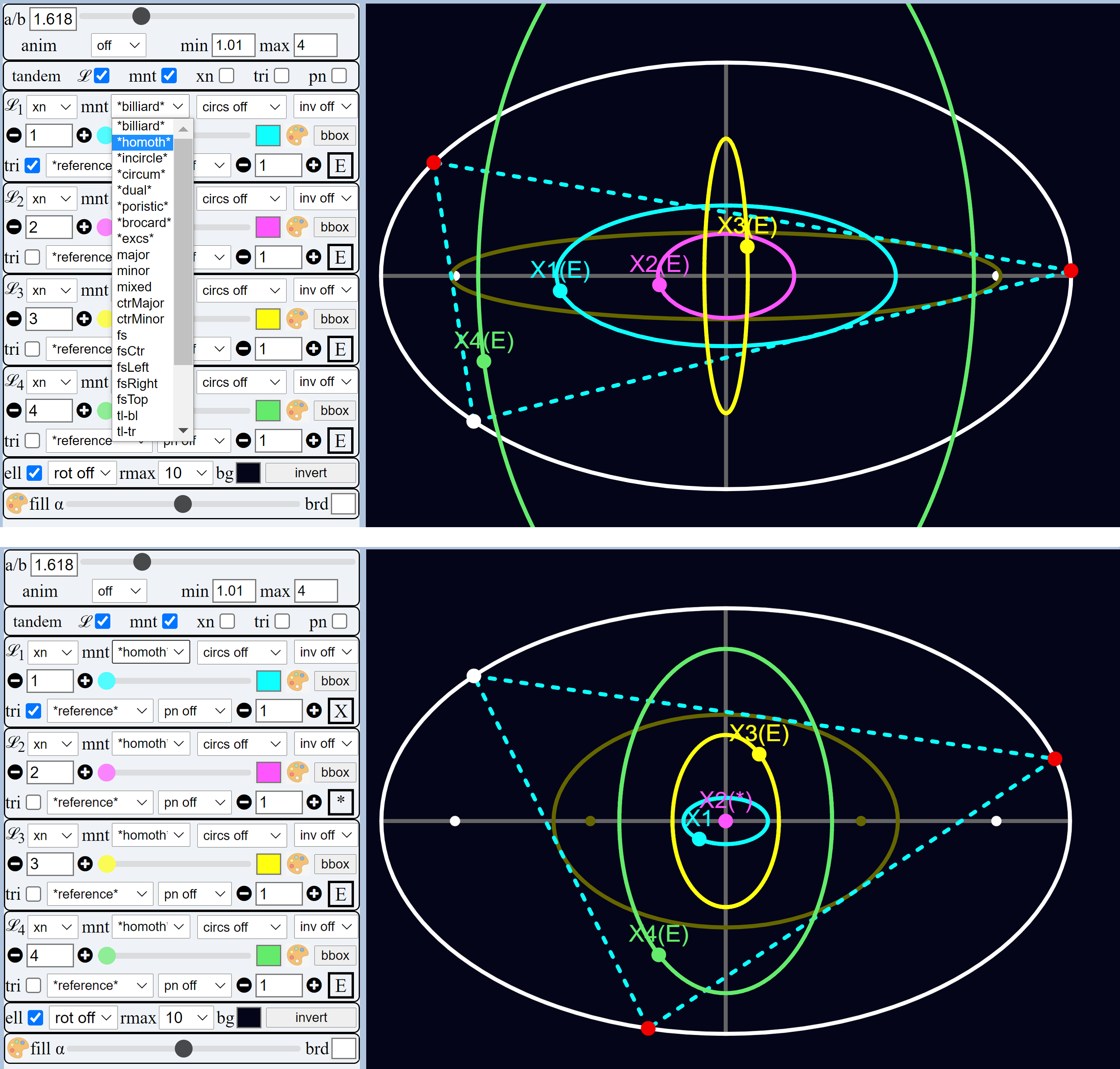}
    \caption{Usage of the tandem feature. The user has previously selected triangle centers $k=1,2,3,4$ for each of the channels. \textbf{Top:} the user is about to flip, in tandem, triangle family from ``billiard'' to ``homothetic'' for the first channel; \textbf{Bottom:} since the tandem \texttt{mnt} is checked, all channels flip in unisn to ``homothetic'', with the visualization being updated in one shot. To quickly flip through all other families, the user can hit the up and down keys on the keyboard.}
    \label{fig:09-tandem-example}
\end{figure}

As an example, consider the sequence shown in \cref{fig:09-tandem-example}. Tandem checkboxes $\Lm$ and \texttt{mnt} are checked, indicating both locus type and triangle family are in unison across all channels. This automatically sets \texttt{xn} across all channels, i.e., loci will be drawn (as opposed to, e.g., the envelope). The use then needs to manually set the triangle center values of 1,2,3,4 for each channel. To now observe these across all families, since \texttt{mnt} is set, the user simply needs to flip through the triangle families using the triangle family drop-down on any one of the channels in the strip. In fact this can be done with the up and down arrows on the keyboard once that control comes into focus (e.g., by expanding the drop-down), allowing for very quick perusal of this phenomenon across all triangle families.

\section{Odds \& Ends}

\subsection{Ellipse, Locus Range, and Animation Background} 

\begin{figure}
    \centering
    \includegraphics[width=.8\textwidth]{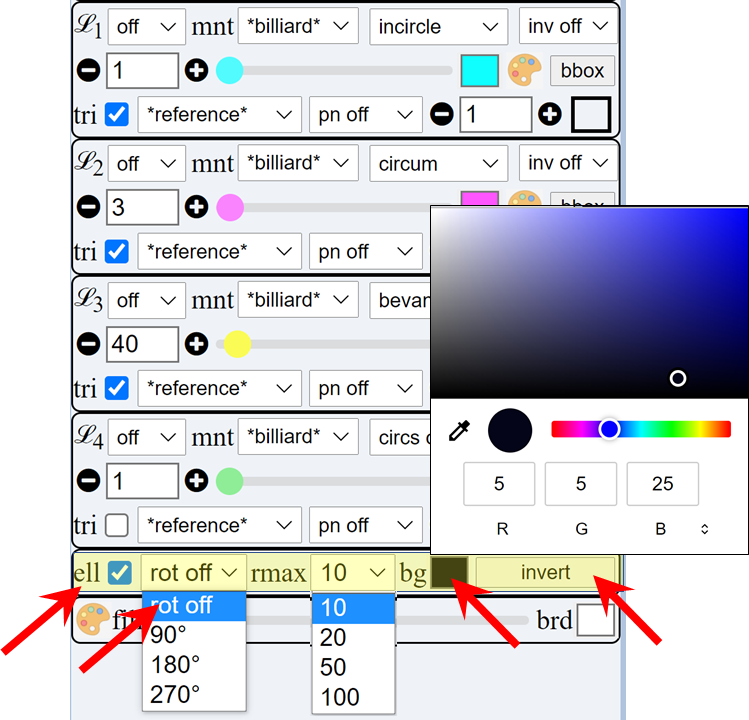}
    \caption{In the highlighted area controls are available to (i) \texttt{ell}: show or hide the main ellipse, (ii) \texttt{rot xxx}: apply a global rotation to the animation widow, (iii) \texttt{rmax}: set the bounding box of the area in which loci are computed, (iv) \texttt{bg}: set the animation window's background color, and (v) \texttt{invert}: invert (RGB negative) colors of background and all loci drawn.}
    \label{fig:09-ellipse-range}
\end{figure}

As shown in  \cref{fig:09-ellipse-range}, the area immediately below the four sets channel controls the following parameters:

\begin{itemize}
\item \texttt{ell} checkbox: whether the main ellipse underlying a triangle family of choice should be drawn or not.
\item Rotation menu: the default \texttt{rot off} setting leaves the animation window as is. Settings  \texttt{$90^\circ$,$180^\circ$,$270^\circ$} apply a global rotation to the picture drawn.
\item \texttt{rmax} menu: the (half-side) of the square bounding box where points in all loci are evaluated, respective to the minor semiaxis of the ellipse, assume to be of unit length. Ideally, this should be set to as small a value as able to contain all loci. 
\item \texttt{bg}: used to set the background color of the main animation window, dark blue by default. By clicking on the colored square an RGB picker window pops-up permitting fine control of the color.
\item \texttt{invert} button: single-click inversion (in RGB space) of colors of background and loci currently in being drawn.
\end{itemize}

\subsection{Resetting the UI and Recentering the Animation} 

\cref{fig:09-anim-appearance} highlights \texttt{reset UI} and \texttt{center UI} push-buttons are located at the top-left corner of the app. These are used to (i) restore all controls in the app to their default values, and (ii) recenter the geometry drawn to the center of the animation, respectively. 

\subsection{Setting the Locus Color}

\begin{figure}
    \centering
    \includegraphics[width=.8\textwidth]{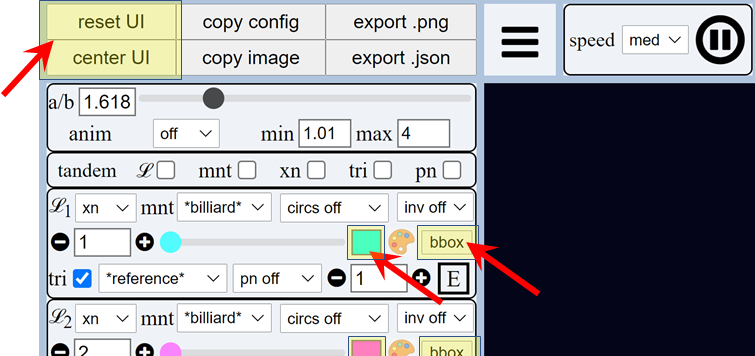}
    \caption{Reset and center push-buttons are located at the top-left corner of the app.  which (i) \texttt{reset UI}: all controls in the app are restored to their default values; (ii) \texttt{center UI}: the center of the simulation is panned back to the center of the animation area. This is useful after having previously panned the picture via a mouse drag. Also shown is (iii) a color selector square located to the right of every triangle center scrollbar, through which a new color can be selected for displaying the corresponding locus. Finally, (iv) a \texttt{bbox} push button is provided to repositing and scale the geometric scene so as to best fit it in the available space.}
    \label{fig:09-anim-appearance}
\end{figure}

Also shown in \cref{fig:09-anim-appearance} are color and rescaling controls to the right of the triangle center scrollbar. These are permit (i) selection of a color specific to a particular locus being drawn, and (ii) a resizing/recentering of the particular locus so as to best fit the animation window.

\subsection{Collapsing the Locus Control Area}

\begin{figure}
    \centering
    \includegraphics[trim=0 0 150 0,clip,width=.6\textwidth]{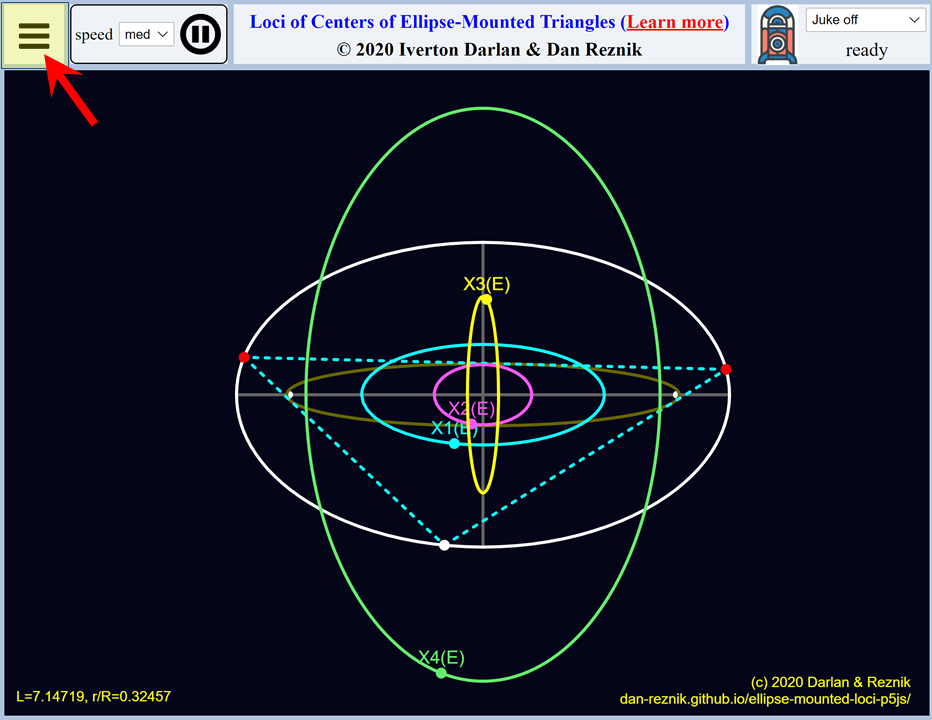}
    \caption{The hamburger control (three horizontal bars) located to the right of the the main controls can be clicked to hides/expand the main controls.}
    \label{fig:09-hamburger}
\end{figure}

The ``hamburger'' control shown in \cref{fig:09-hamburger} can be used to hide/expand the set of controls on the left marging of the app, sometimes useful for demonstration purposes.

\section{Arsty Color Fills}

\begin{figure}
    \centering
    \includegraphics[width=.8\textwidth]{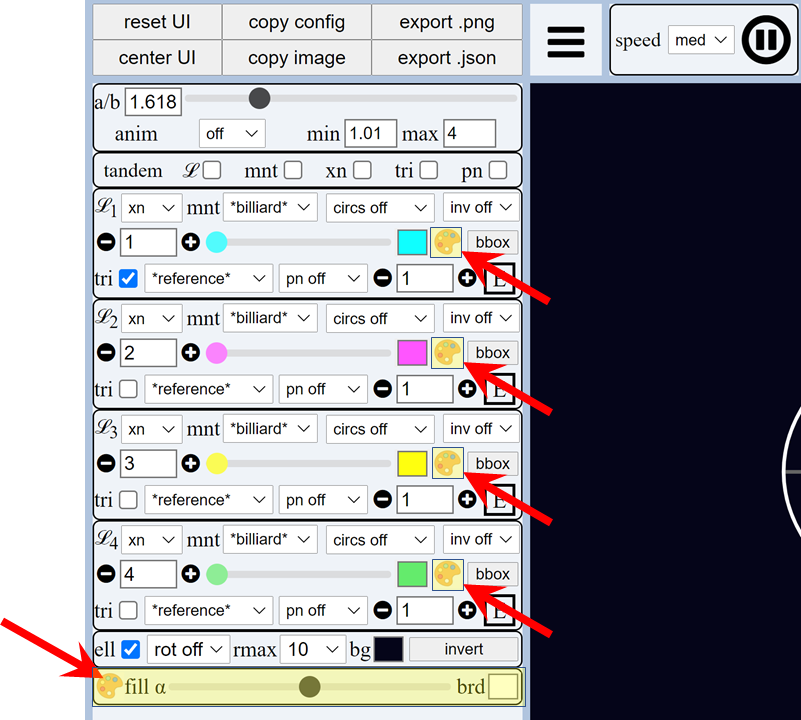}
    \caption{Clicking on the highlighted palette icons in the mid-right section of every channel control area triggers color fills in any drawn loci. Clicking it several times (resp. right clicking on it) randomizes colors (resp. removes the color fills). At the bottom of the channel control strip a scrollbar can be used to control the transparency of the fills. A color chooser at its right side can be clicked to select the color of region borders (default is white).}
    \label{fig:09-color-filling}
\end{figure}

A set of controls, highlighted in \cref{fig:09-color-filling}, can be used to color fill connected regions of loci. A first clicking on the palette icon in the middle-right section of a channel's control group selects a random set of pastel colors. Subsequent clicks (or hitting the right arrow key) generate a new random color set. Hitting the left arrow goes back to color sets  previously generated. Right-clicking on the palette icon and or changing any other setting in the user interface causes the color fills to disappear.

Also highlighted in \cref{fig:09-color-filling} is a scrollbar and color chooser located below the bottom-most channel control group. These are used to set (i) 
the transparency of colors fills, and (ii) the color of the border of connected regions (default is white).

A collage of four colored-filled curvaceous loci is shown in \cref{fig:09-showcase}. Some two hundred such ``artsy'' loci are showcased in \cite{reznik2021-artful}.

\section{Sharing and Exporting}

As shown in \cref{fig:09-sharing-exporting}, four buttons on the global control strip (top left of the interface) can be used to copy and export a link, an image, or a vector graphics representation the loci currently rendered. The buttons are as follows:

\begin{figure}
    \centering
    \includegraphics[width=.7\textwidth]{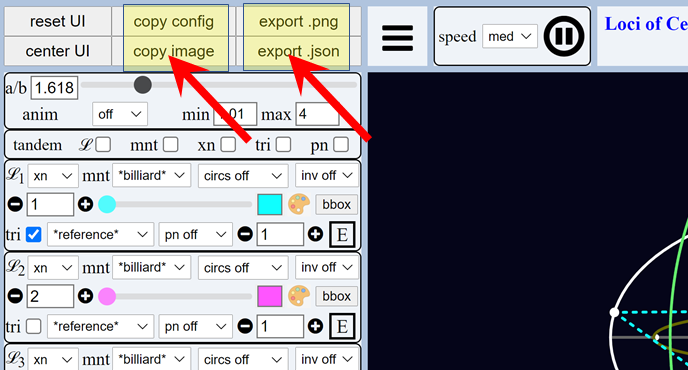}
    \caption{Buttons on the upper control strip for copying, sharing, and exporting experiment configuration and images.}
    \label{fig:09-sharing-exporting}
\end{figure}

\begin{itemize}
\item \texttt{copy config}: A URL containing all information pertaining to the current geometric scene (as defined by the channel controls and other pieces of UI) is copied to the clipboard. This URL can be shared with another user and/or shortened prior to sharing, e.g., with \texttt{bit.ly}.
\item \texttt{copy image}: the image currently on the animation window is copied to the clipboard. It can then be pasted anywhere else as an image.
\item \texttt{export .png}: what is currently on the animation window is downloaded to the local file system.
\item \texttt{export .json}: a vector graphics representation of all loci drawn is exported in human-readable JSON format.
\end{itemize}

\section{Jukebox Playback}
\label{sec:juke}

Many experiments constructed with the tool are stored in a database accessible by the app. These are organized in different thematic groups which can be played back in continuously in ``jukebox'' mode (each experiment being displayed 5-10 seconds). This is initiated by selecting a thematic group from the drop-down at the top right hand side of the app window, highlighted in \cref{fig:09-juke}. To quickly zip forward or backward thru items in the series, click (or right-click) on the jukebox icon to the left of the drop-down. Jukebox mode can be stopped at anytime by returning the drop-down to the \texttt{Juke off} setting.

Currently, upwards of 450 experiments illustrating many phenonema manifested by Poncelet triangle families (PCFs) have been organized in 20+ playlists, divided into the following 3 groups: (i) aesthetically-oriented, (ii) research-oriented observations, and (iii) exercise and conjecture ``truckloads'' of exercises. Intringuinly, most phenomena in (ii) and (iii) lack proofs.

\subsection{Aesthetic}

\begin{itemize}
\item \texttt{01 single line} (48 items): early experiments with single loci drawn as plain curves
\item \texttt{02 colored lines} (14 items): multiple loci each drawn with a different color
\item \texttt{03 dark mode} (57 items): multiple loci with dark background
\item \texttt{04 region filled} (74 items): multiple loci with their interior filled with various colors for artistic effect
 \end{itemize}
 
 \subsection{Research}
 
 \begin{itemize} 
\item \texttt{05 envelopes} (4 items): exploring envelopes of PCFs
\item \texttt{06 cevians} (3 items): exploring loci of PCF cevians triangles
\item \texttt{07 inversives} (11 items): loci and phenomena of PCF inversive images wrt various circles
\item \texttt{08 constant power} (11 items): PCF configurations whereby certain points maintain constant power with respect to some notable circle
\item \texttt{09 poncelet} (9 items): various conjectures around PCFs and derivatives
\item \texttt{10 polar} (11 items): loci and phenomena of polar images of PCFs
\item \texttt{11 graves} (15 items): phenomena involving the ``Graves'' triangle of various PCFs
\item \texttt{17 subtris} (16 items): phenomena involving $X_k$-subtriangles (triple subdivision)
\item \texttt{18 concurring normals} (5 items): illustration of PCFs with concurring normals
\item \texttt{19 focal invs and cevs} (10 items): inversive images and cevians of PCFs wrt to foci
\item \texttt{20 mixtilinears} (7 items): phenomena involving mixtilinear circles
\item \texttt{21 flank tris \& outer extouches} (13 items): loci and phenomena involving external triad triangles
\item \texttt{22 polar poncelet and astables} (6 items): illustration of astable loci (point + curve) of certain PCFs
\item \texttt{23 subcevians and subanticevians} (8 items): loci of triangle centers and vertices of sub (triangulated) cevians and anticevians
\item \texttt{24 bicentric tangential and orthic} (8 items): locus of the so-named triangles over the bicentric family
\item \texttt{25 circumcimson \& bitangs} (11 items): experiments with the circumcimson and various bitangent triangles over the bicentric and incircle families
   
\end{itemize}

\subsection{Exercise and conjecture ``truckloads''}

A set of four playlists involving a long lists of experiments still lacking proof are proposed in:

\begin{itemize}
\item \texttt{12 truckload I: pole-polars} (35 items): phenomena involving poles and polars of PCFs
\item \texttt{13 truckload II: billiard cevs \& peds} (20 items): cevian and pedals deriving from the confocal (billiard) family
\item \texttt{14 truckload III: incircle cevs \& peds} (30 items): cevian and pedal phenomena deriving from various PCFs
\item \texttt{15 truckload IV: homoth cevs \& peds} (19 items): cevian and pedal phenomena deriving from the homothetic family
\item \texttt{16 truckload V: more polars \& peds} (7 items): more polar and pedal phenomena deriving from the homothetic family
\end{itemize}

%% file: 910_app_poncelet.tex
\section{Poncelet's Porism}
\label{app:poncelet}

\begin{figure}
    \centering
    \includegraphics[width=\textwidth]{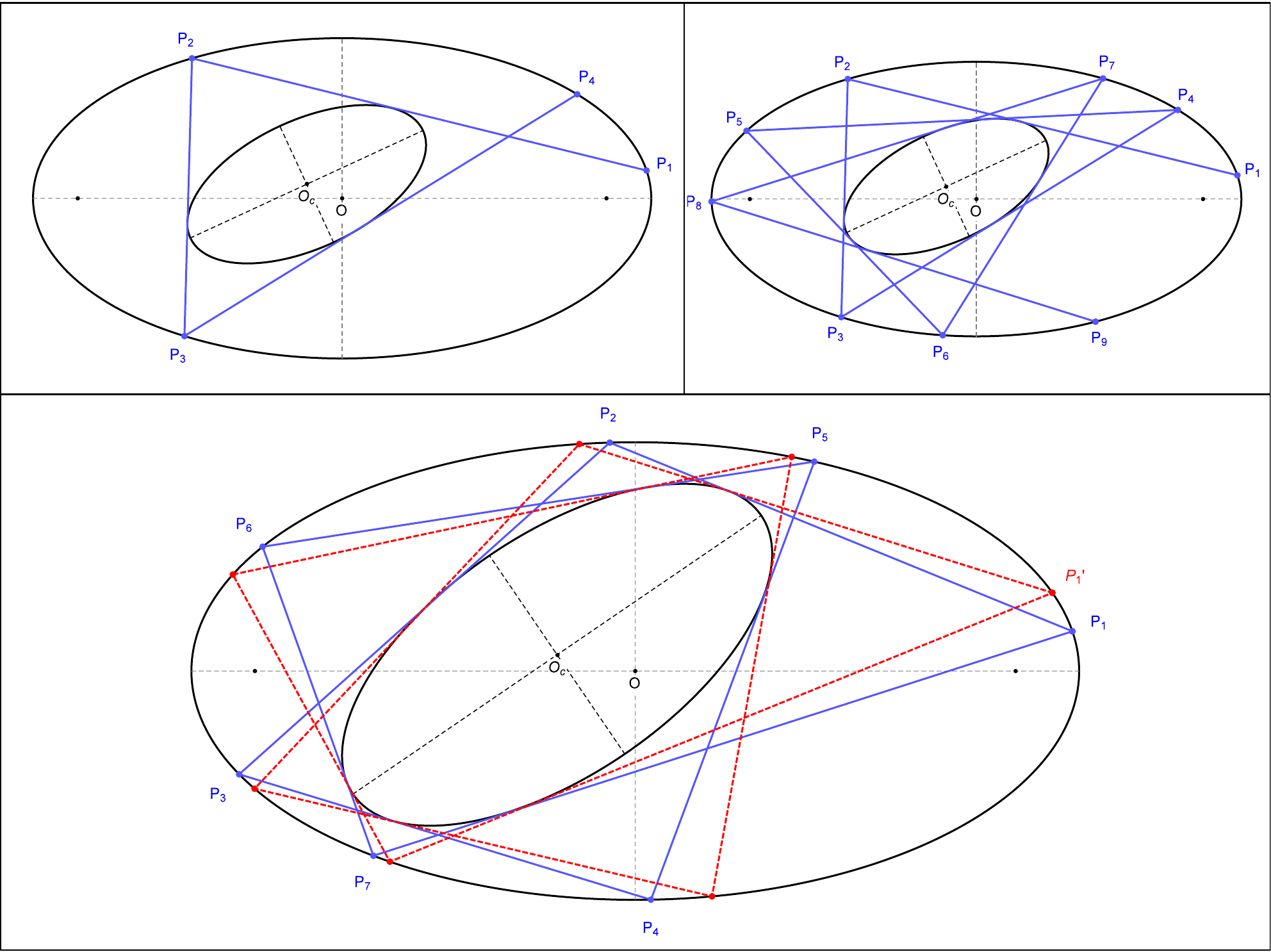}
    \caption{\textbf{Top left}: 3 Poncelet iterations within a pair of ellipses in general position; their centers are labeled $O$ and $O_c$, respectively. \text{Top right:} 5 more iterations executed (starting at $P_4$), showing the trajectory is not likely to close. \textbf{Bottom:} a new ellipse pair for which an iteration departing from $P_1$ closes after 7 steps (blue polygon). Poncelet's porism guarantees that if the iteration were to start anywhere else on the outer ellipse, e.g., $P_1'$, it will also yield a closed, 7-gon (dashed red). \href{https://youtu.be/kzxf7ZgJ5Hw}{Video},  \href{https://observablehq.com/@dan-reznik/poncelet-iteration}{Live}}
    \label{fig:poncelet}
\end{figure}

Given two real conics $\E$ and $\E'$, a {\em Poncelet transverse} is constructed as follows (see \cref{fig:poncelet}): pick a point $P_1$ on $\E$ and shoot a ray in the direction of one of the tangents to $\E'$. Let $P_2$ denote the non-$P_1$ intersection of this ray with $\E$, i.e., $P_1 P_2$ is a chord of $\E$. Now shoot a ray from $P_2$ in the direction of a tangent to $\E'$ and find $P_3$, etc., repeating this a number of times. For certain conic pairs, the transverse closes after $N$ steps, i.e., $P_N=P_1$. Pairs with this property are said to satisfy ``Cayley's conditions'', see \cite{dragovic11} details.

Poncelet's closure theorem (PCT) states that if such an $N$-step transverse closes, than one that starts from any other point on $\E$ will also close in $N$ steps, i.e., one has a 1d family of Polygons \cite{centina15}. 

In the specific case where $\E$ and $\E'$  are concentric, axis-aligned ellipses, the Cayley condition for such a pair to admit a family of triangles reduces to   \cite{georgiev2012-poncelet}:

\[ \frac{a'}{a} + \frac{b'}{b} = 1 \]
where $a,b$ (resp. $a',b'$) are the semiaxes of $\E$ (resp. $\E'$).

A few such pairs are shown in \cref{fig:six-caps}. These include a pair (i) with a fixed incircle (and external ellipse), (ii) a fixed circumcircle (and internal inellipse), (iii) of homothetic ellipses, (iv) of ``dual'' ellipses (these have reciprocal aspect ratios), and (v) of the {\em excentral} triangles\footnote{The excentral triangle has sides through the vertices of a reference one and along the external bisectors.} to the confocal pair itself.